\documentclass[superscriptaddress,twocolumn,pre]{revtex4}
\usepackage{amsmath}
\usepackage{amsfonts}
\usepackage{amssymb}
\usepackage{graphicx}
\graphicspath{{images/}} 
\usepackage{color}
\usepackage[pdfstartview=FitH,
            breaklinks=true,
            bookmarksopen=false,
            bookmarksnumbered=true,
            colorlinks=true,
            linkcolor=black,
            citecolor=black,
            urlcolor=black,
            pdftitle={Transitions in optimal adaptation strategies in fluctuating environments},
            pdfauthor={AM, TM, OR, and AW},
            pdfsubject={Paper}
            ]{hyperref}

\def\(({\left(}
\def\)){\right)}
\def\[[{\left[}
\def\]]{\right]}
\newcommand{\beq}{\begin{equation}}
\newcommand{\eeq}{\end{equation}}
\newcommand{\ceqref}{\eqref}
\newcommand{\<}{\langle}
\renewcommand{\>}{\rangle}
\newcommand{\B}{\boldsymbol}
\newcommand{\ud}{\mathrm{d}}
\newcommand{\changed}[1]{#1}

\DeclareMathOperator*{\argmax}{arg\,max}

\usepackage{mathtools}
\DeclarePairedDelimiter\bra{\langle}{\rvert}
\DeclarePairedDelimiter\ket{\lvert}{\rangle}
\DeclarePairedDelimiterX\braket[2]{\langle}{\rangle}{#1 \delimsize\vert #2}

\begin{document}

\title{
    Transitions in optimal adaptive strategies for populations in fluctuating environments
}

\author{Andreas Mayer} 
\affiliation{Laboratoire de physique th\'eorique,
    CNRS, UPMC and \'Ecole normale sup\'erieure,
    75005 Paris, France}
\author{Thierry Mora}
\affiliation{Laboratoire de physique statistique,
    CNRS, UPMC and \'Ecole normale sup\'erieure,
    75005 Paris, France}
\author{Olivier Rivoire} 
\affiliation{Center for Interdisciplinary Research in Biology,
    CNRS, INSERM and Coll\`ege de France,
    75005 Paris, France}
\author{Aleksandra M. Walczak}
\affiliation{Laboratoire de physique th\'eorique,
    CNRS, UPMC and \'Ecole normale sup\'erieure,
    75005 Paris, France}

\address{}

\date{\today}

\begin{abstract}
Biological populations are subject to fluctuating environmental conditions. Different adaptive strategies can allow them to cope with these fluctuations: specialization to one particular environmental condition, adoption of a generalist phenotype that compromise between conditions, or population-wise diversification (bet-hedging).
Which strategy provides the largest selective advantage in the long run depends on the range of accessible phenotypes and the statistics of the environmental fluctuations. Here, we analyze this problem in a simple mathematical model of population growth. First, we review and extend a graphical method to identify the nature of the optimal strategy when the environmental fluctuations are uncorrelated. Temporal correlations in environmental fluctuations open up new strategies that rely on memory but are mathematically challenging to study:  we present here new analytical results to address this challenge. We illustrate our general approach by analyzing optimal adaptive strategies in the presence of trade-offs that constrain the range of accessible phenotypes. Our results extend several previous studies and have applications to a variety of biological phenomena, from antibiotic resistance in bacteria to immune responses in vertebrates.

\end{abstract}

\maketitle

\section{Introduction}

Nothing is as constant as change. This age-old adage applies to biological populations, which may respond by evolving mechanisms to mitigate the consequences of environmental fluctuations \cite{Levins1968,Seger1987,Kussell2005,Chevin2010,Simons2011}. This adaptation can be implemented at different levels. At an individual level, the simplest strategy consists in adopting a generalist phenotype that \changed{does} reasonably well across environments. At a population-level, another strategy is to constantly generate a phenotypically diverse mixture of individuals, each specialized to a different environmental condition.
Which strategy provides the largest selective advantage in the long run depends on the nature of environmental fluctuations and on the fitness costs and trade-offs limiting the range of accessible phenotypes. For instance, although tracking the environment to adopt a phenotype specialized to each current condition may seem optimal, this strategy is often precluded by the costs of constantly monitoring environmental changes and of frequently switching between phenotypes.

Which strategies to deal with environmental fluctuations may be selected is a long-standing question in evolutionary biology.
Interest in this question has recently been rekindled by novel laboratory experiments with populations growing in controlled fluctuating environments \cite{Acar2008,Beaumont2009,Lambert2015}, new theoretical developments providing links to ideas from information theory and stochastic thermodynamics \cite{Vinkler2014,Kobayashi2015a,Rivoire2016}, and its relevance to understanding non-genetic modes of inheritance \cite{Lachmann1996,Rivoire2014} and how biological populations might respond to climate change \cite{Chevin2010,Botero2015}.

Here, we study this question in a model of population growth in a \changed{randomly} fluctuating environment. The model considers a large population of organisms characterized by their phenotype and replicating at discrete generations. An optimal adaptive strategy is defined by the choice of phenotypes and switching rates between them that ensures the largest long-term population growth rate. We analyze how this optimal strategy depends on the environmental statistics and the replication rates. The analysis reveals transitions between qualitatively different strategies: non-switching or single-phenotype strategies, where all of the population is of the same phenotype; and switching or bet-hedging strategies, where the population diversifies. Further transitions arise between strategies where the population adopts a phenotype specialized in a single environment, and strategies relying on a generalist phenotype.

Our work extends the growing literature investigating transitions between optimal adaptive strategies \cite{Donaldson-Matasci2008,Salathe2009,Rivoire2014,Patra2015,Skanata2016} and generalizes some of our previous results on the adaptation of immune strategies to pathogen statistics \cite{Mayer2016}. In particular, we derive exact expressions for the transitions between different modes of immunity in memoryless environments when the strategy includes an adjustable investment into immunity. We also calculate analytically the transitions between switching and non-switching strategies between two phenotypes in temporally correlated environments.
After briefly introducing the mathematical framework (Sec.~\ref{secsetup}), we present a graphical method for studying transitions in optimal adaptive strategies in temporally uncorrelated environments (Sec.~\ref{seciid}), and apply it to the case of an immune system with adjustable investment (Sec.~\ref{secprotection}).
We then turn to the case of temporally correlated environments and provide new analytical and numerical results on transitions in this more general setting (Sec.~\ref{seccorrelated}).

\section{Population growth in fluctuating environments}
\label{secsetup}

\begin{figure*}
    \includegraphics{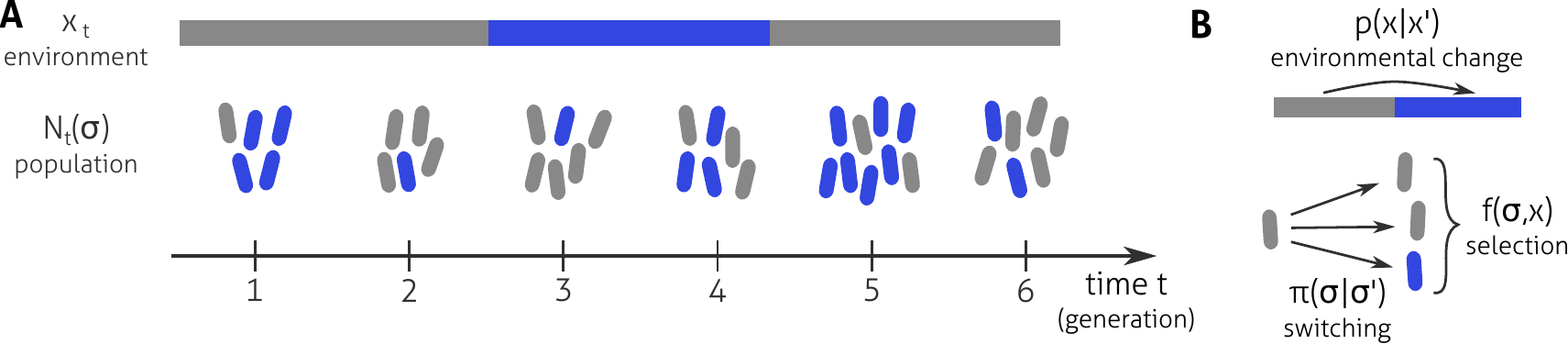}
    \caption{{\bf Model of population growth in a fluctuating environment.} (A) A population composed of individuals of different phenotypes $\sigma$ grows in a changing environment $x_t$. Between each discrete generation, the phenotype of each individual may switch. (B) The environment follows a stochastic dynamics described by a Markov chain with transition rates $p(x|x')$. The population composition changes between generations due to the effects of selection (an individual with phenotype $\sigma$ in environment $x$ produces in average $f(\sigma, x)$ offspring) and phenotype switching (an individual with phenotype $\sigma'$ has probability $\pi(\sigma|\sigma')$ to have an offspring with phenotype $\sigma$).
        \label{figsetup}}
\end{figure*}

We are interested in describing the evolution of a possibly phenotypically heterogeneous biological population (of cells, organisms, etc.) in a fluctuating environment.
We describe the population at generation $t$ by the number $N_t(\sigma)$ of individuals with a given phenotype $\sigma$.
Phenotypes differ by their replication rate $f(\sigma, x)$, which give the mean number of offspring  produced by an individual of phenotype $\sigma$ in environmental condition $x$ (see Fig.~\ref{figsetup}).
The environment is described as a discrete Markov chain with a transition matrix $p(x|x')$, which we assume to be stationary and ergodic.
The population changes under the influence of the selective pressures generated by the differences in replication rates between phenotypes, and through phenotype switches described by a transition matrix $\pi(\sigma|\sigma')$.
In the limit of infinitely large population size, the population composition follows the recursion \cite{Rivoire2011}
\begin{equation} \label{eqNrecursion}
    N_{t+1}(\sigma) = f(\sigma, x_t) \sum_{\sigma'} \pi(\sigma|\sigma') N_t(\sigma').
\end{equation}
This equation can also be written in a compact matrix notation as
\begin{equation} \label{eqNrecursionvectorial}
    \B N_{t+1} = \B A^{(x_t)} \B N_t, \quad \text{with} \quad A_{\sigma, \sigma'}^{(x_t)} = f(\sigma, x_t) \pi(\sigma|\sigma').
\end{equation}
Here and in the following, we write vectors and matrices in bold notation. 

The different modalities by which populations might cope with fluctuating environmental conditions correspond to different properties of the switching matrix $\pi(\sigma|\sigma')$.
For \textit{non-switching strategies}, the whole population has the same phenotype $\tilde \sigma$ and the switching matrix consists in a row of ones, $\pi(\sigma | \sigma') =1$ if \changed{$\sigma=\tilde\sigma$} and 0 otherwise.
If the chosen phenotype is a better all-rounder doing intermediately well across environments, this corresponds to an individual-level \textit{generalist strategy}.
For \textit{switching strategies}, we may distinguish those with and without memory.
In a switching strategy without memory, the probability of switching to a phenotype does not depend on the parental phenotype, $\pi(\sigma|\sigma') = \pi(\sigma)$.
Such strategies \changed{implement
population-level bet-hedging}, i.e., diversification of the population into phenotypes that may each be specialized to one of the environmental conditions to come.
Switching with memory, where $\pi(\sigma|\sigma')$ does depend on $\sigma'$, provides the basic ingredients, variation and heritability, to enable adaptive tracking of the environment through Darwinian evolution. In the limit where switching is very rare, $\pi(\sigma|\sigma') \ll \pi(\sigma' | \sigma')$ for $\sigma\neq\sigma'$, the phenotypic dynamics is equivalent to the strong-selection weak-mutation limit of population genetics \cite{Rivoire2014}. The model thus integrates in a common mathematical framework a range of different modes of response to environmental variations.

Over long evolutionary time scales, selection might act on the adaptive mechanisms to adjust them to the statistics of environmental fluctuations.
Explicit models of the evolution of the switching rates $\pi(\sigma|\sigma')$ show that variation in switching rates can indeed be selected upon \cite{Salathe2009,Carja2014,Rivoire2014}. Transgenerational feedback reinforcing the production of successfull phenotypes provides an alternative mechanism to learn a good strategy \cite{Xue2016}.
Which adaptive strategy do we expect to evolve in the long run?
Here, we focus on the optimal strategy representing the optimal possible end-product of this evolution. In our model, the optimal switching rates maximize long-term growth rate, defined as
\begin{equation} \label{eqLambdadef}
    \Lambda = \lim_{T\rightarrow\infty} \frac{1}{T} \ln N_T/N_0,
\end{equation}
where $N_T = \sum_\sigma N_T(\sigma)$ is the total population size.
To understand why this is the relevant measure of evolutionary success in the long run, consider a population with two subpopulations following different strategies.
Then in the long run the population following the strategy with highest long-term growth rate almost surely outnumbers the one following the other strategy for almost every sequence of environments \cite{Cover2005}. 
The question of which adaptive strategy $\pi^*(\sigma|\sigma')$ has the largest selective advantage is thus recast as the problem of maximizing the long-term growth rate over possible strategies: 
\begin{equation} \label{eqoptproblem}
\pi^*(\sigma|\sigma') =\argmax_{\pi(\sigma|\sigma')} \Lambda,
\end{equation}
for given replication rates $f(\sigma, x)$ and given environmental dynamics $p(x|x')$. This is the problem that we address in this paper.

\section{When and how to be a generalist in uncorrelated environments}
\label{seciid}

\subsection{Extended fitness set and Pareto optimality}
\begin{figure*}
    \includegraphics{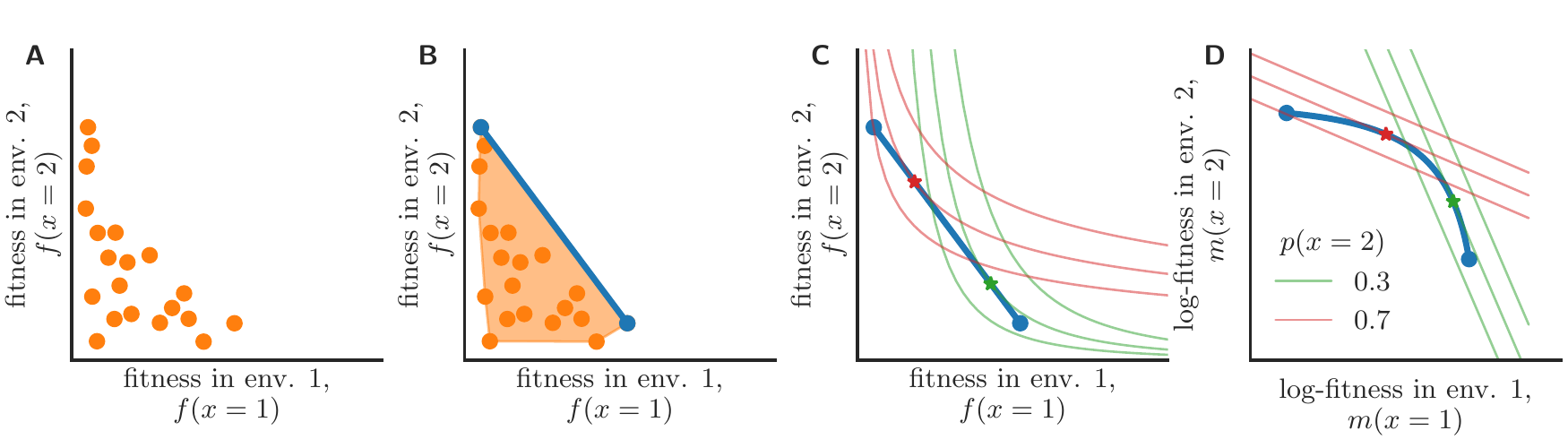}
    \caption{{\bf Illustration of the steps of a graphical method of finding the best adaptation strategy in uncorrelated environments.} (A) Fitness values of phenotypes across environments (orange dots). (B) Fitness values achievable by switching strategies (orange area) are those inside the convex hull of the fitness values of the different phenotypes. A necessary condition for optimality is to lie on the Pareto frontier (blue line). (C, D) The optimal strategy has the fitnesses (red/green star) at which the isolines of the long-term growth rate for given environmental frequencies (red lines for $p(2)=0.7$, green lines for $p(2)=0.3$) are tangential to the Pareto frontier. (C) In fitness space the isolines are curved. (D) To determine the optimal strategy it is more convenient to work in log-fitness space, where the isolines are straight lines.
        \label{figgraphical}}
\end{figure*}

The simplest environmental fluctuations to consider are memoryless fluctuations, where the state of the environment is independent of its state in the previous generation, $p(x|x') = p(x)$.
In this case, no gain can be expected from keeping a memory of past phenotypic states, and the optimal adaptive strategy is also memoryless, $\pi(\sigma | \sigma') = \pi(\sigma)$.
Since the population composition is constant over generations, the number of offspring depends only on the state of the environment and
\ceqref{eqNrecursion} reduces to a recursion for the total population size $N_t=\sum_\sigma N_t(\sigma)$:
\begin{equation} \label{eqNtotrecursion}
    N_{t+1} =  
N_t f(x_t),
\end{equation}
where
\begin{equation} \label{eqfx}
    f(x) = \sum_\sigma f(\sigma, x) \pi(\sigma)
\end{equation}
is the average population fitness.
Graphically, it is convenient to represent each possible phenotype $\sigma$ as a point in the space of environmental conditions $x$ (where each environment $x$ defines a dimension), with coordinates given by the replication rates $f(\sigma,x)$ (orange dots in Fig.~\ref{figgraphical}A). The set $D_f = \{\sum_\sigma f(\sigma, x) \pi(\sigma) | \sum_\sigma \pi(\sigma) = 1, \pi(\sigma) \geq 0 \}$ of achievable $f(x)$ when switching rates $\pi(\sigma)$ are varied then corresponds to the convex hull of these points (orange area in Fig.~\ref{figgraphical}B). In the ecological literature, this set of achievable strategies is known as the extended fitness set and was introduced by Levins \cite{Levins1968}.

The recursion for the total population size \eqref{eqNtotrecursion} is solved by $N_T = N_0 \prod_t f(x_t)$.
Taking logarithms, we have $\ln N_T/N_0 = \sum_{t=1}^T \ln f(x_t)$ and we can apply the law of large numbers to write the long-term growth rate \eqref{eqLambdadef} as a weighted average of log-fitnesses:
\begin{equation} \label{eqLambdalog}
    \Lambda = \sum_x p(x) \ln f(x),
\end{equation}
with weights given by the frequency of each environment.

Finding the optimal strategy $\pi^*(\sigma)$ that maximizes $\Lambda= \sum_x p(x) \ln \sum_\sigma f(\sigma, x) \pi(\sigma)$ over the domain allowed by the rules of probabilities
is a convex optimization problem whose solution is well known
 \cite{Cohen1966,Levins1968,Haccou1995,Cover2005,Rivoire2011,Donaldson-Matasci2008}. 
It is useful to rephrase the problem as the optimization of $\Lambda=\sum_x p(x)\ln f(x)$ over the fitnesses $\B f$ constrained to belong to the extended fitness set $D_f$ introduced above. One can go further and equivalently optimize $\Lambda=\sum_x p(x)m(x)$ over the log-fitnesses $m(x)=\ln f(x)$ contrained to belong to $\ln (D_f)$.
Going from $\B \pi$ to $\B f$ to $\B m$ simplifies the expression of the objective function $\Lambda$ but makes the domain of optimization more complex.

Eq.~\eqref{eqLambdalog} shows that the long-term growth rate is an increasing function of each environment fitness $f(x)$. Increasing fitness in one environment is always desirable if this can be done without impairing fitness in any other environment. 
Thus, any optimal solution must lie on the set of fitnesses $\B f$ for which no improvement can be made in one environment without impairing performance in another, called the Pareto frontier.
Usually, no phenotype provides the best fitness for all environments due to trade-offs between performance under different conditions. Thus
the Pareto frontier is generally not a single point but a line when the environment alternates between two conditions (blue line in Fig.~\ref{figgraphical}B), and a hyper-surface of dimension $n-1$ when the environment alternates between $n$ conditions.
To find the overall optimum along the Pareto front requires to consider the explicit way in which performances for different objectives combine into a scalar measure, which is determined in our case by the frequency of the different environments  \eqref{eqLambdalog}.

\subsection{Graphical method for finding the optimal strategy}

The various views of the optimization problem discussed in the previous subsection imply a graphical method to determine the optimal strategy. For simplicity, we illustrate it by considering switching between only two environments (Fig.~\ref{figgraphical}). Starting from the graphical representation of the Pareto front for the set of achievable fitnesses (Fig.~\ref{figgraphical}B), we need to find the point of this frontier with the highest growth rate: this is done graphically by representing the growth rate isolines $\Lambda[f(1),f(2)] = K$ (red and green lines in Fig.~\ref{figgraphical}C where the two colors corresponds to different environmental statistics)
given by \eqref{eqLambdalog}:
\begin{equation}
f(2) = \frac{e^{K/p(2)}}{f(1)^{p(1)/p(2)}}.
\end{equation}
By plotting the isolines for different $K$ we can find the isoline for the largest $K$ that still intersects with the Pareto frontier, called supporting line.
The intersection point defines the optimal adaptive strategy the population should adopt (red and green stars in Fig.~\ref{figgraphical}C).
This construction was first proposed by Levins \cite{Levins1968}.

Here, we propose to go one step further and work in log-fitness space to circumvent the difficulty of handling curved isolines. In log-fitness space, the isolines are linear and normal to the vector $\B p$:
\begin{equation} \label{eqiso}
    p(1) m(1) + p(2) m(2) = K.
\end{equation}
If the Pareto front has a tangent of slope $-p(1)/p(2)$, the tangent point thus defines the optimal strategy for the environment $\B p$ (Fig.~\ref{figgraphical}D).
More generally, the supporting isoline corresponding to the optimal growth rate shares at least one point with the Pareto frontier but is otherwise entirely above that frontier.

The graphical method generalizes to $d$ environments by studying the extended fitness set in a space of $d$ dimension, according to the following procedure.
First, represent the phenotypes' fitnesses as points in the space of different environments, each environment defining a dimension (orange dots in Fig.~\ref{figgraphical}A).
Second, construct the convex hull of these points to find the fitnesses achievable by switching strategies $D_f$ (orange area in Fig.~\ref{figgraphical}B), and find the Pareto-optimal frontier of that set (blue line in Fig.~\ref{figgraphical}B).
Third, plot this Pareto surface in log-fitness space (blue line in Fig.~\ref{figgraphical}D).
Finally, find the hyperplane normal to $\B p$ that is a supporting hyperplane of the Pareto frontier (red and green lines in Fig.~\ref{figgraphical}D), and
read off the optimal strategy as the intersection point between that hyperplane and the Pareto frontier (red and green stars in Fig.~\ref{figgraphical}D).

When the Pareto frontier is contained in a hyperplane, fitnesses can be rescaled onto the unit simplex, $\sum_x f(x)=1$, with no loss of generality \cite{Donaldson-Matasci2008}. In this case the optimal strategy is given in terms of the rescaled fitnesses as $\B f^*=\B p$, making the graphical construction even simpler (Fig.~\ref{figunitsimplex} and App.~\ref{appmapping}).

\subsection{Transitions between switching, non-switching, and generalist strategies}
\label{seciidtransitions}

\begin{figure*}
    \includegraphics{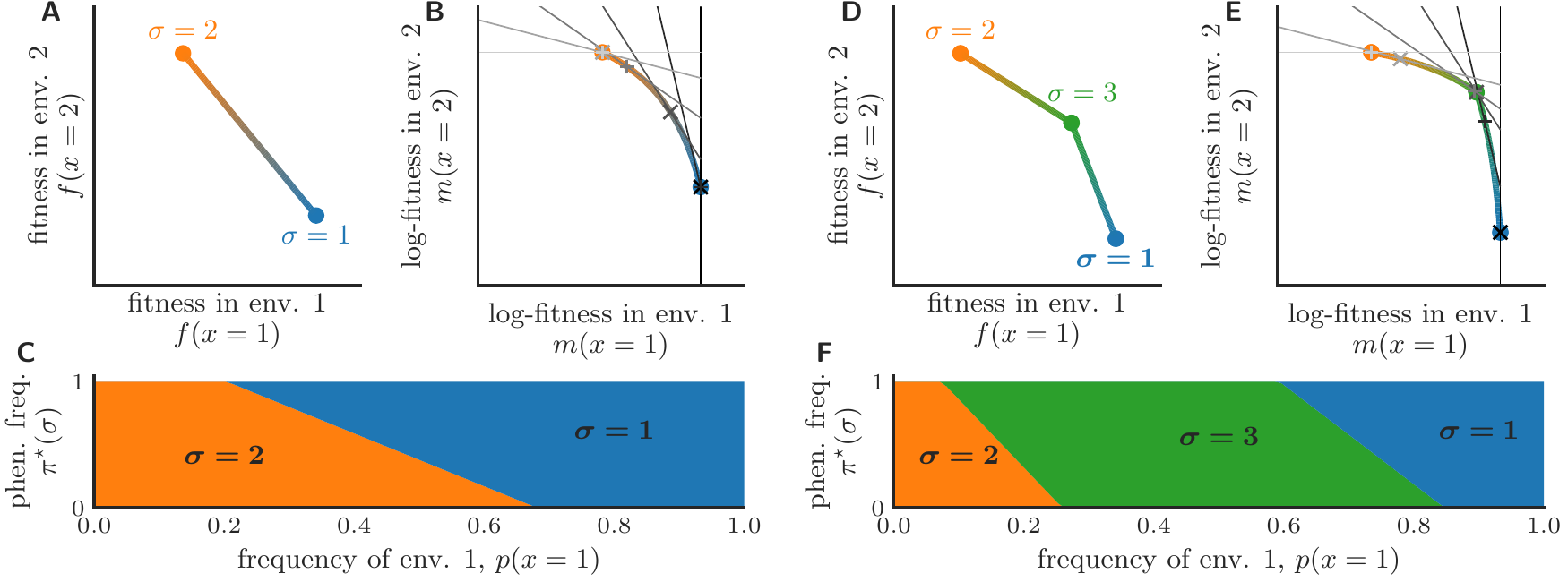}
    \caption{{\bf Transitions of the optimal strategy as a function of environmental frequencies without (A-C) and with (D-F) a generalist phenotype.} 
        (A,D) Pareto frontier of achievable fitness vectors by phenotypes (dots) and their mixtures (lines).
        (B,E) In log-fitness space a tangent construction (grey lines) yields the optimal strategy (grey crosses) for different environments (from dark to light grey for $p_1=1 \to p_1 = 0$ in $0.2$ steps).
        (C,F) Transitions between switching and non-switching strategies as a function of the probability of encountering environment 1.
        \changed{Parameters: (A-C) $f(\sigma=1) = (1, 0.3), f(\sigma=2) = (0.4, 1)$, (D-F) $f(\sigma=1) = (1, 0.2), f(\sigma=2) = (0.3, 1.0), f(\sigma=3)=(0.8, 0.7)$.}
        \label{figtransitions}}
\end{figure*}

The graphical method provides a visual approach to classify the different possible adaptive strategies.
For the sake of simplicity, we start again with the case of a two-state environment and first assume that only two phenotypes are accessible: a blue phenotype ($\sigma=1$) best suited to environment 1 and an orange phenotype ($\sigma=2$) best suited to environment 2 (Fig.~\ref{figtransitions}A-C). In this case, the Pareto front is a segment joining the two phenotypes. In log-fitness space, this segment is curved and concave, implying that $\partial m(2)/\partial m(1)$ is a decreasing function of $m(1)$. Different environmental statistics are characterized by the frequencies $p(1)$ and $p(2)=1-p(1)$ of the two environmental states. The value of $p(1)$ sets the slope $-p(1)/p(2)$ of the isolines of growth rate that we should consider \ceqref{eqiso}. 

Depending on the value of $p(1)$, different cases arise. First, if $p(1)$ is too high or too low, there is no tangent to the Pareto front of slope $-p(1)/p(2)$ and the support point lies at one of the two extremities of the Pareto front. In these cases, the optimal strategy (crosses in Fig.~\ref{figtransitions}B) is to adopt a constant phenotype -- the phenotype optimal for the most frequent environmental state. When $p(1)$ takes an intermediate value, the isoline is tangent to the Pareto frontier at an intermediate support point, indicating an optimal strategy involving switching between the two possible phenotypes. As a function of the 
frequency of encountering different environments, there are thus two transitions, from non-switching to switching and to non-switching again. This succession of optimal strategies is read off as a function of the environmental frequency from the Pareto line (Fig.~\ref{figtransitions}C). 

One can make the problem more interesting by adding
a third ``generalist'' phenotype, which does relatively well across both environments (Fig.~\ref{figtransitions}D-F, green dot). This generalist creates a kink in the Pareto frontier, meaning that it will be optimal as a constant phenotype for a certain range of environmental conditions.
Thus, depending on the frequencies of the two environmental states, the optimal strategy consists either of having a constant specialized phenotypes (blue or orange) when one environment is much more frequent than the other, a constant generalist phenotype (green) when the two environments have similar frequencies, or switching between a specialized phenotype and the generalist phenotype in intermediate situations (Fig.~\ref{figtransitions}F). The transition from specialist to generalist was studied in a similar model in \cite{Donaldson-Matasci2008}, but in the slightly different context of a continuous choice of strategies.

These conclusions generalize to an arbitrary number $d$ of environmental states.
It follows from the graphical construction that for a given statistics of the environment, the number of discrete phenotypes between which the population may switch in optimal strategies is at most equal to the number of different environmental conditions, $d$: the subset of the extended fitness set corresponding to this switching is the polytope of dimension $d-1$ whose vertices are these $d$ phenotypes (a segment for $d=2$, a triangle for $d=3$). This observation may be viewed as extending to changing environments the principle of competitive exclusion stating that a single niche cannot support more than one species.

\begin{figure*}
    \begin{center}
    \includegraphics{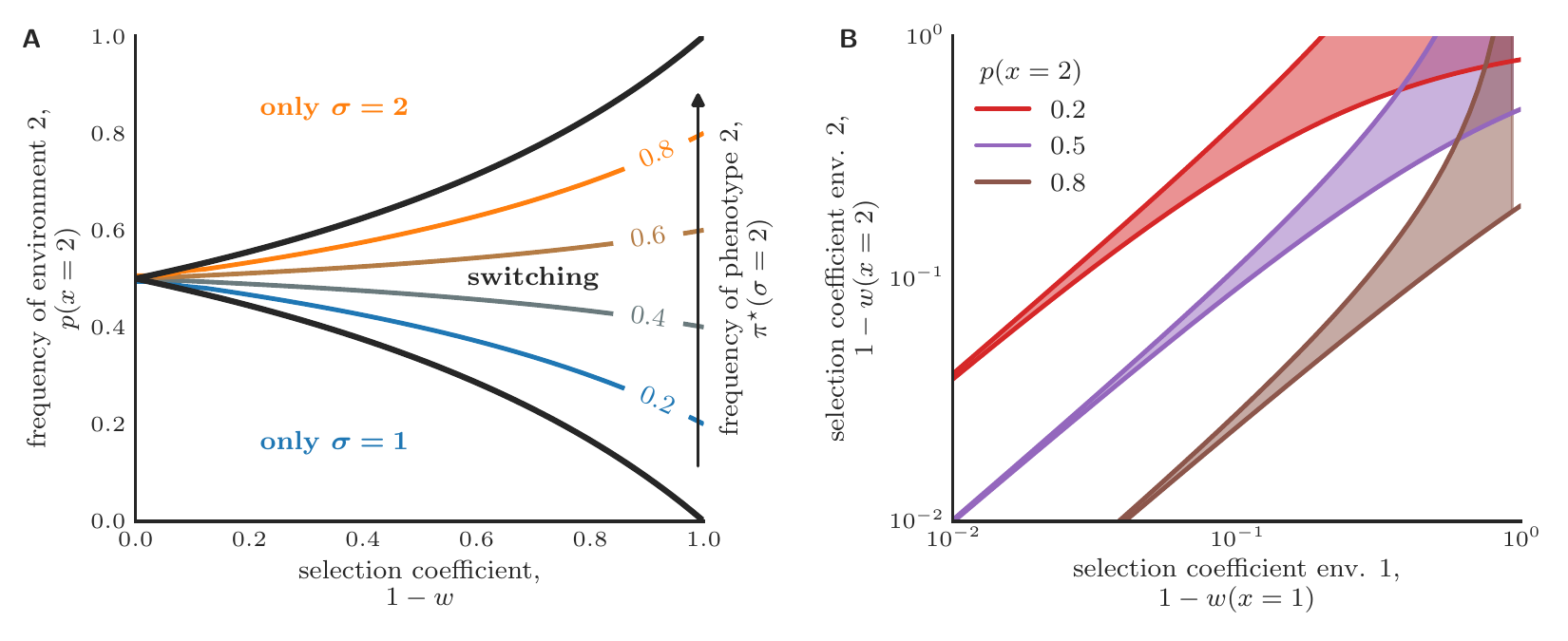}
        \caption{{\bf Transitions between \changed{switching and non-switching} strategies depend on environmental selectivity and environmental frequencies.}
        In a temporally uncorrelated environment changing randomly between two states, $1$ and $2$, a population of organisms is adapted optimally by either being in a \changed{single} phenotypic state or by having a mixture of phenotypes (bet-hedging) depending on the statistics of the environment and the degree to which the phenotypes are specialized.
        In environment $x=1$ ($2$) phenotype $2$ ($1$) \changed{has replication rate $w_x$} relative to the other phenotype.
        (A) Transitions as a function of specialization level and environmental frequency in the symmetric case, \changed{$w_1 = w_2 = w$}. The black lines mark the transition from single-phenotype to bet-hedging strategies: above the upper (lower) line the entire population optimally has phenotype 2 (1), between the two lines phenotypic diversification provides an advantage.
        The optimal fraction of phenotype $2$ in the bet-hedging region is shown by the colored lines.
        (B) Regions of selection factors in which bet-hedging is the preferred strategy (shaded areas) for environments with different frequencies of being in state $2$. Either strong selection or a precise mapping between the relative selection factors and the relative environmental frequencies are needed to make bet-hedging optimal. 
      \label{figiidbethedging}}
    \end{center}
\end{figure*}

We complement the graphical analysis by analytical results in the simplest case of two environments and two phenotypes illustrated by Fig.~\ref{figiidbethedging}A-C. Since only the relative fitnesses in each environment is relevant for the dynamics, we set without restriction of generality the replication rate of each phenotype in its preferred environment to 1. The other phenotype has a selective disadvantage, with replication rate $w_x<1$:
\begin{equation} \label{eqselectioncoefficients}
    f(\sigma, x) = \begin{cases} 1 & \sigma = x,\\ w_x & \sigma \neq x.\end{cases}
\end{equation}
The parameter $w_x$ can be interpreted as the degree of specialization: $w_x=1$ means no specialization, while $w_x=0$ means extreme specialization.
Since $p(1) = 1 - p(2)$ and $\pi(1) = 1-\pi(2)$, there are just two free parameters $p_2\equiv p(2)$ and $\pi_2\equiv \pi(2)$.
In these variables the long-term growth rate \eqref{eqLambdalog} is written as
\begin{equation} \label{eqLambdainnate}
    \begin{split}
    \Lambda = \, &p_2 \log [(1-\pi_2)w_2+\pi_2] \\
        &+ (1-p_2)\log[1-\pi_2+\pi_2w_1)].
    \end{split}
\end{equation}
To find the optimal fraction of the population with phenotype 2, $\pi_2^\star$, Eq.~\eqref{eqLambdainnate} is to be maximized over $\pi_2 \in [0, 1]$.
The optimization yields
\begin{equation} \label{eqpioptinnate}
\pi_2^\star = \left\{\begin{array}{cl}
    0  & \textrm{if}\quad p_2 \leq p_2^{\rm lb},\\ 
    \frac{p_2 - p_2^{\rm lb}}{p_2^{\rm ub} - p_2^{\rm lb}} &\textrm{if}\quad p_2^{\rm lb} < p_2 < p_2^{\rm ub}, \\
    1 &\textrm{if}\quad p_2 \geq p_2^{\rm ub}\\ 
\end{array}\right.
\end{equation}
with lower and upper bounds
\begin{align} \label{eqplb}
    p_2^{\rm lb} = \frac{w_2}{1 + (1-w_2)w_1/(1-w_1)} \\
    p_2^{\rm ub} = \frac{1}{1 + (1-w_2)w_1/(1-w_1)}
\end{align}
on the environmental frequencies for which diversification is optimal.
The calculation recapitulates the conclusions from the graphical method (Fig.~\ref{figiidbethedging}C).
In the limit were selection is very stringent, $w_1 \rightarrow 0$ and $w_2 \rightarrow 0$, the transitions disappear, $p_2^{\rm lb} \to 0$ and $p_2^{\rm ub} \to 1$ and the optimal strategy reduces to proportional betting,
\beq\label{proportionalbetting}
\pi_2^\star = p_2.
\eeq
In the context of biological bet-hedging, this result was already noted by Cohen \cite{Cohen1966}; it was also derived earlier in the context of gambling by Kelly \cite{Cover2005}.

The range of environmental frequencies for which bet-hedging is favored over the non-switching strategies depends strongly on the selectivity of the environments (Fig.~\ref{figiidbethedging}A).
Consider for simplicity the symmetric case $w_1 = w_2 = w$, then non-switching strategies are favored for $\left|p_2 - 1/2\right| \geq (1-w)/[2(1+w)]$.
In the limit, $w \rightarrow 0$, the strategy tends to proportional bet-hedging as discussed earlier.
The larger $w$, the smaller the region of environmental frequencies for which switching strategies are optimal.
As there is smaller variability in fitness across generations for the same phenotype, switching is less needed to hedge against environmental fluctuations.

Instead of considering transitions in optimal strategies as environmental frequencies are varied, we can also consider transitions as selection pressures are varied \changed{at} fixed environmental frequencies (Fig.~\ref{figiidbethedging}B).
As selection pressures are decreased there are transitions to a non-switching strategy (white areas in Fig.~\ref{figiidbethedging}B).
The optimality of bet-hedging (shaded areas in Fig.~\ref{figiidbethedging}B) for weak selection pressures depends on a precise matching between the asymmetry in selection pressures and environmental frequencies.
This conclusion generalizes the results of \cite{Salathe2009}, which considered numerically asymmetric fitness landscapes, $w_1 \neq w_2$, but only with a symmetric environment, $p_1 = p_2 = 1/2$.


\section{Transitions between optimal immune strategies}
\label{secprotection}
\changed{Fitnesses achievable by single phenotypes (orange dots in Fig.~\ref{figgraphical}) can fill a set delimited by a continuous line, called {\em trade-off function}, which is the Pareto frontier of non-switching strategies.
It is common to consider such a continuous set of phenotypes with all possible \changed{switching strategies} between them \cite{Levins1968,Donaldson-Matasci2008}. The Pareto frontier of switching strategies defined in the previous section then delimits the convex hull of that continuous set.
The two Pareto frontiers (of switching and non-switching strategies) coincide if the trade-off function is concave, i.e. if the set of achievable phenotypes is convex; in that case non-switching strategies are optimal everywhere.
Otherwise, similar transitions as in the previous section will arise \cite{Donaldson-Matasci2008}.
In some biological situations however, only some combinations of phenotypes along a trade-off function may be accessible at the same time, meaning that one cannot switch between all phenotypes on the trade-off line. Such a constraint on switching rates can induce discontinuous transitions, or cause the co-existence of multiple locally optimal solutions, as we now illustrate in a simple model of evolution of immunity.
}

\begin{figure*}
    \begin{center}
        \includegraphics[width=\textwidth]{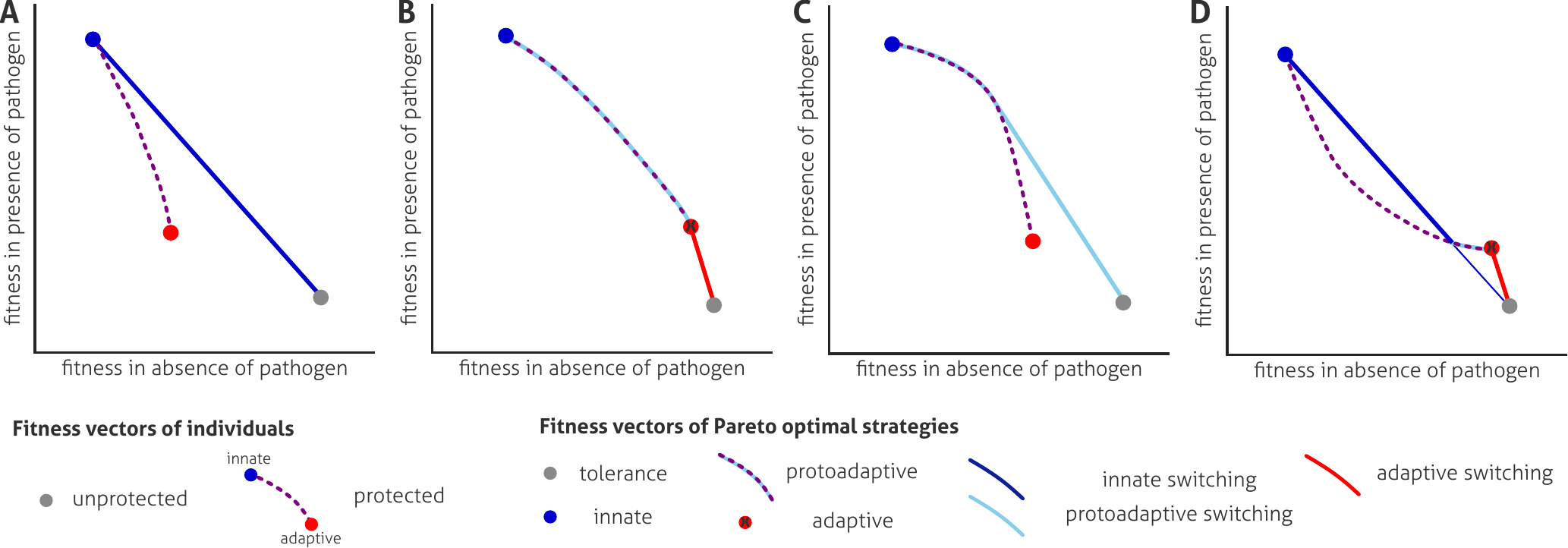}
        \caption{{\bf Strength of trade-offs between constitutive and defense cost of protection determine adaptation strategy in a fluctuating pathogenic environment.}
        In the model, unprotected individuals have a fixed fitness profile (grey dot).
        Protection comes in various degrees of adaptability (dashed purple line) between maximal (blue dot) and minimal (red dot) level of constitutive investement in defense.
        Switching strategies are possible where only parts of the population are protected. They have fitnesses that are a linear combination of the fitness of unprotected and protected indviduals for a given level of adaptability.
        The optimal strategy needs to lie along the Pareto frontier of the possible fitnesses.
        The strategies that lie on the Pareto surface allow reading off the succession of optimal strategies as the probability of encountering the pathogen is decreased.  
        (A) Strong trade-offs lead to switching strategies being better then adaptable protection.
        (B) For shallow trade-offs the Pareto frontier is achieved by adaptable defenses.
        (C) A combination of shallow and steep trade-offs can lead to only some degree of adaptability being used. 
        (D) A concave trade-off function can lead to first order transitions in strategy and potential co-existence of locally optimal solutions.
    \label{figimmune}}
    \end{center}
\end{figure*}

Our illustrative example is a model that we proposed to explain the diversity of immune strategies observed across the tree of life \cite{Mayer2016}. The purpose is to show how different strategies are associated with different statistics of pathogen dynamics. In its simplest form, the model has two environmental states, presence ($x=1$) or absence ($x=0$) of a pathogen. In a given strategy, it has two accessible phenotypes, protected ($\sigma=1$) or unprotected ($\sigma=0$).
 Strategies are represented by $\B f=(f(x=0),f(x=1))$ as before.

The unprotected phenotype is fixed in fitness space: $\B f=(f_{\rm base},f_{\rm inf})$  (grey dot in Fig.~\ref{figimmune}), where $f_{\rm inf}<f_{\rm base}$ is the reduced fitness in infected  unprotected individuals. By constrast, the protected phenotype lies on a trade-off function: $\B f =(f_{\rm con}, f_{\rm def}(f_{\rm con}))$, with $f_{\rm con} \in [f_{\rm con}^{\rm min}, f_{\rm con}^{\rm max}]$ (dashed purple line delimited by red and blue dots in Fig.~\ref{figimmune}). $f_{\rm con}<f_{\rm base}$ represents the reduced fitness due to the investment into the protection, while $f_{\rm def}>f_{\rm inf}$ is the fitness of protected individuals in presence of the pathogen.

The choice of $f_{\rm con}$ along the trade-off function sets the investment into the protection, and is part of the strategy: once this strategy is fixed, it is possible to switch between protected and unprotected phenotypes, but not between different points of the trade-off function. This constraint can be justified biologically by the high cost of plasticity that such switches would incur.

The function $f_{\rm def}(f_{\rm con})$ encodes the trade-off between the efficiency of the protection and its cost. By analogy with immune mechanisms in vertebrates, we interpret it in terms of {\it adaptivity} of the response within the lifetime of the organism, with higher adaptivity enabling lower cost at the expense of lower protective efficiency \cite{Mayer2016}. We therefore refer to the maximally protective and costly strategy with $f_{\rm con}=f_{\rm con}^{\rm max}$ as {\it innate immunity} and to the minimally protective and costly strategy with $f_{\rm con}=f_{\rm con}^{\rm min}$ as {\it adaptive immunity}. Intermediate strategies with $f_{\rm con}^{\rm min}<f_{\rm con}<f_{\rm con}^{\rm max}$ are referred to as {\it protoadaptive}. 

Within this model, the equation for long-term growth rate in an uncorrelated environment \eqref{eqLambdalog} becomes
\begin{equation} \label{eqLambdadirect}
    \begin{split}
        \Lambda = &p \ln[\pi f_{\rm def} + (1-\pi) f_{\rm inf}] \\&+ (1-p) \ln [\pi f_{\rm con} + (1-\pi) f_{\rm base}],
    \end{split}
\end{equation}
where $p\equiv p(x=1)$ is the probability of the presence of the pathogen and $\pi\equiv \pi(x=1)$ the probability of being protected. Here, the problem is not only to find the optimal switching probability $\pi^\star$, but also to find the optimal protection adaptability, $f_{\rm con}^\star$. To summarize, the problem is as follows: for a given $p$, $f_{\rm inf}, f_{\rm base}$ and $f_{\rm def}(f_{\rm con})$, find $f_{\rm con}^\star$ and $\pi^\star$ that maximize long-term growth rate in Eq.~\ref{eqLambdadirect}.

We are particularly interested in transitions between $f_{\rm con}^\star, \pi^\star$ taking intermediate or extremal values within their respective ranges. Given that each of these two variables can either reach its lower or upper bound or take an intermediate value, nine different cases may arise. However, since the level of adaptability of the response is inconsequential if none of the population is protected ($\pi^\star = 0$), only seven qualitatively different immune defense strategies are relevant: tolerance ($\pi^\star = 0$, grey dot in Fig.~\ref{figimmune}), innate ($\pi^\star = 1$, $f_{\rm con} = f_{\rm con}^{\rm max}$, blue dot in Fig.~\ref{figimmune}), adaptive ($\pi^\star = 1$, $f_{\rm con} = f_{\rm con}^{\rm min}$, red crossed dot in Fig.~\ref{figimmune}), protoadaptive ($\pi^\star = 1$, $f_{\rm con}^{\rm min}<f_{\rm con}<f_{\rm con}^{\rm max}$, light blue line with purple dashes in Fig.~\ref{figimmune}), innate switching ($0<\pi^\star<1$, $f_{\rm con} = f_{\rm con}^{\rm max}$, blue line in Fig.~\ref{figimmune}), adaptive switching ($0<\pi^\star<1$, $f_{\rm con} = f_{\rm con}^{\rm min}$, red line in Fig.~\ref{figimmune}), and protoadaptive switching ($0<\pi^\star<1$, $f_{\rm con}^{\rm min}<f_{\rm con}<f_{\rm con}^{\rm max}$, light-blue line in Fig.~\ref{figimmune}).

Which of these strategies is optimal in a given environment? 
And what is the nature of the transitions between strategies as the frequency of encountering the pathogen is varied?
Here, we apply the graphical method to answer these questions and show how the answers depend critically on the shape of the trade-off function. Our conclusions, summarized in Fig.~\ref{figimmune}, are supported by analytical results derived in Appendix~\ref{appiid}.

The simplest case is when adaptability comes at an excessive cost, as depicted in Fig.~\ref{figimmune}A: an innate switching strategy is then always preferable to an adaptive strategy. In this case, as the probability of encountering the pathogen increases, the optimal strategy transitions from tolerance (grey dot in Fig.~\ref{figimmune}A) to an innate defense strategy (blue dot in Fig.~\ref{figimmune}A) via an innate switching (blue line in Fig.~\ref{figimmune}A).
When adaptability of the defense does not impair its effectiveness as severely, as in Fig.~\ref{figimmune}B, two new transitions occur. As the probability of encountering the pathogen increases, the optimal strategy now transitions from tolerance to, successively, adaptive switching, adaptive, protoadaptive and finally innate defense strategy. In other cases, a switching protoadaptive defense strategy may also be optimal, as in the case of the trade-off function of Fig.~\ref{figimmune}C.
In this case, as the probability of encountering the pathogen increases, the optimal strategy transitions from tolerance to, successively, protoadaptive switching, protoadaptive and finally innate defense strategy.
Finally, we may consider a case where the trade-off line is not convex as in Fig.~\ref{figimmune}D.
The Pareto frontier is then not necessarily concave, and we might have first order transitions between strategies.
For the trade-off shape shown in Fig.~\ref{figimmune}D, there is a transition from protoadaptive switching (blue line with purple dashes) directly to innate switching (blue line), with a discontinuity in the level of adaptability of the response.

\section{When and how to use memory in temporally correlated environments}
\label{seccorrelated}

In temporally correlated environments, the past phenotypes of an individual carry information about the next environmental state. The optimal switching strategy may thus involve memory, i.e. it may be advantageous for $\pi(\sigma|\sigma')$ to depend on $\sigma'$. 
Stochastic switching with memory serves an additional purpose relative to the memoryless switching strategies considered so far: in addition to providing a bet-hedging mechanism against the uncertainty of the environment, it provides the variation and heritability needed for tracking the environmental state.
Here, we extend the previous analysis to characterize the conditions under which temporal correlations in environmental fluctuations favor switching strategies with memory over non-switching strategies. The graphical method does not extend to correlated environments but we show that the transitions between switching and non-switching strategies can be characterized analytically.

\subsection{Insights from the adiabatic limit}
\label{secadiabatic}

It is instructive to start with long correlation times, when the duration of each environmental state is much longer than the time that it takes for the population to reach its steady state composition. In this adiabatic limit, the model is analytically solvable \cite{Kussell2005,Rivoire2011}.
We present the solution for the case where switching takes place between a number of different phenotypes, with each phenotype $\sigma$ being best in one environment $x$, which we denote by the same symbol $\sigma = x$ (other cases can in fact always be reduced to this one \cite{Rivoire2011}). 
A calculation based on a series of eigendecompositions of the growth matrix in different environments (see Appendix~\ref{appadiabatic} for derivation) leads to an expression of the long-term growth rate as \cite{Rivoire2011}
\begin{equation} \label{eqadiabatic}
    \begin{split}
    \Lambda = &\sum_x p(x) \ln f(x, x)\\
              &+ \sum_{x, x'} p(x|x') p(x') \ln[\pi(x|x') \Gamma(x, x')],
    \end{split}
\end{equation}
which involves the overlap $\Gamma(x, x')$ between steady-state population compositions in environments $x, x'$, given by 
\begin{equation} \label{eqGamma}
    \Gamma(x, x') =
        \frac{f(x, x')}{f(x', x') - f(x, x')} + \frac{f(x, x)}{f(x, x) - f(x', x)},
\end{equation}
if the environment changes, $x \neq x'$, and $1$ otherwise.

Optimizing \ceqref{eqadiabatic} over $\pi(x|x')$ subject to the normalization constraint leads to $\pi^\star(x|x') = p(x|x')$.
Within the adiabatic limit, the optimal strategy is therefore always to diversify, with switching rates equal to the environmental switching rates.
This generalizes the result that proportional betting is optimal in the limit of strong selection, \ceqref{proportionalbetting}, to the case where reaching steady state takes longer but environmental switches are rarer. In contrast to the results in the previous section, \changed{switching} is always favored in the adiabatic limit, even when selection is \changed{weak}.

We can use the expression of \ceqref{eqadiabatic} to ask how much each phenotype $\sigma$ should be specialized to its environement $x=\sigma$. Being more specialized means higher fitnesses of the adapted phenotypes, $f(x,x)$, at the expense of lower fitnesses for the maladapted phenotypes, $f(x,x'\neq x)$, assuming a trade-off between the two. 
More specialized phenotypes have lower \changed{relative replication rate} $w(x, x') = f(x, x') / f(x', x')$ [$w(x,x')$ reduces to $w_{x'}$ of \ceqref{eqselectioncoefficients} in the case of two environmental states]. Rewriting
\begin{equation} \label{eqGammarel}
    \Gamma(x, x') = \frac{w(x, x')}{1 - w(x, x')} + \frac{1}{1 - w(x', x)},
\end{equation}
we see that specialization also implies lower overlaps $\Gamma(x,x')$, and thus lower values for the second term in the long-term growth rate \ceqref{eqadiabatic}. On the other hand, the first term in \ceqref{eqadiabatic} grows with $f(x,x)$, i.e. with higher specialization. Because of these contradictory terms, \changed{the optimal strategy along the trade-off between $f(x,x)$ and $f(x,x')$ will depend on the details of trade-off function and of the environmental statistics.} However, as environment fluctuations become slower, $p(x|x'\neq x)\to 0$, the second term in \ceqref{eqadiabatic} vanishes for $x\neq x'$, letting the first term dominate. In that limit, highly specialized phenotypes become more and more advantageous. This observation is again in contrast with the results of the preceding section (Fig.~\ref{figtransitions}D-F), which have shown that generalists are optimal under certain environmental conditions.

\subsection{Connecting the limit of uncorrelated and adiabatically switching environments numerically}

\begin{figure}
    \begin{center}
    \includegraphics{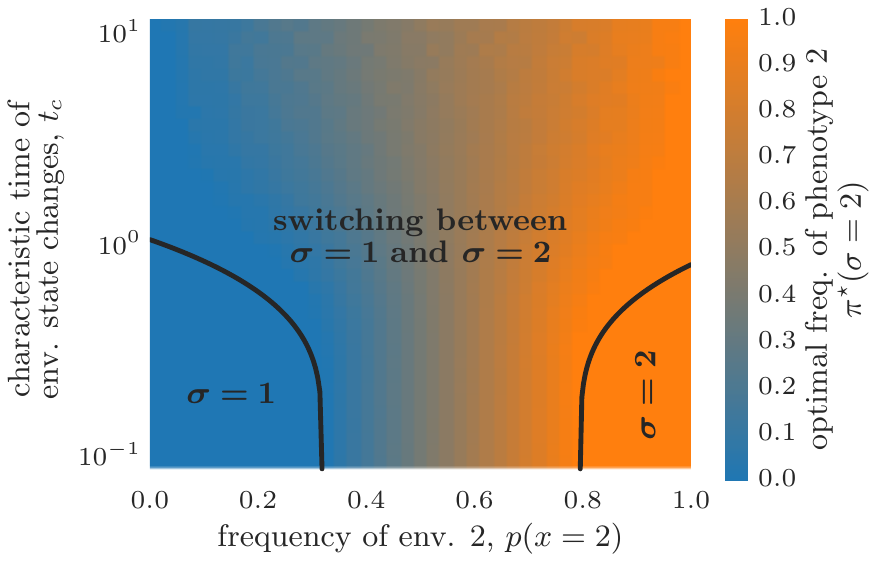}
        \caption{{\bf Switching strategies are favored over a larger range of conditions if environmental states are temporally autocorrelated.}
        Here we generalize the results of Fig.~\ref{figtransitions}A-C about transitions between switching and non-switching strategies by considering the influence of environmental correlation.
        The numerically obtained optimal switching rate $\pi^\star(\sigma=2)$ is plotted as a function of $t_c$, the characteristic time scale of environmental changes, and $p(x=2)$, the fraction of the time the environment is in state 2.
        The range of environmental frequencies in which there is switching ($0 < \pi^\star < 1$) increases with temporal correlations.
        As a comparison we also show the analytical transition lines obtained in Sec.~\ref{seccorrana}, Eqs.~\eqref{eqcorrlb}-\eqref{eqcorrub}.
      \label{figphasesinnate}}
    \end{center}
\end{figure}

\begin{figure*}
    \begin{center}
    \includegraphics{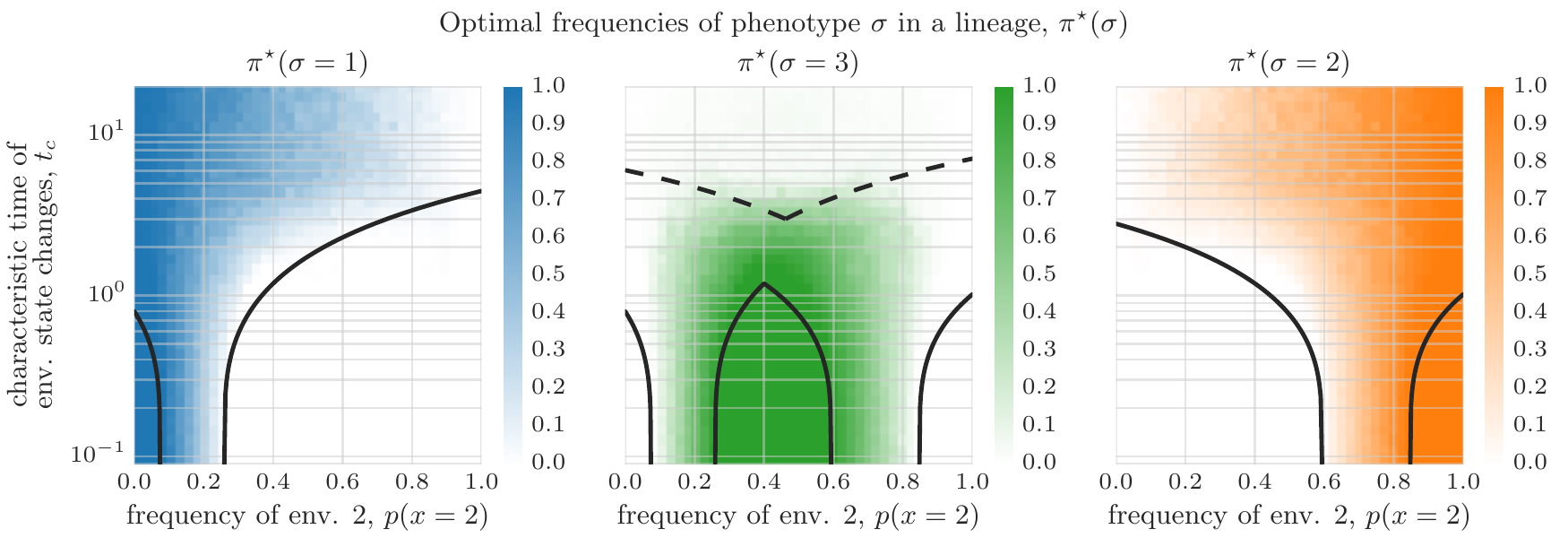}
        \caption{{\bf Switching between specialists is the preferred adaptation strategy in highly correlated environments even if a generalist phenotype is optimal in uncorrelated environments.}
        Here we generalize the results of Fig.~\ref{figtransitions}D-F about transitions between switching, specialist, and \changed{single-phenotype} generalist strategies by considering the influence of environmental correlation.
        The numerically determined optimal frequencies of different phenotypes $\pi^\star(\sigma)$ in a lineage are plotted as a function of $t_c$, the characteristic time scale of environmental changes, and $p(x=2)$, the fraction of the time the environment is in state 2.
        As a comparison we also show the analytical transition lines between single-phenotype and switching strategies obtained in Sec.~\ref{seccorrana}, Eqs.~\eqref{eqcorrlb}-\eqref{eqcorrub} (solid lines) and the approximate transition line above which switching takes place between the two specialist phenotypes as obtained in Sec.~\ref{secadiabatic} and App.~\ref{appadiabatic}, Eq.~\eqref{eqtcspecialists} (dashed lines).
      \label{fig3state}}
    \end{center}
\end{figure*}

So far we have considered two opposite limits: temporally uncorrelated environments in Sec.~\ref{seciid} and \ref{secprotection}, and temporally correlated environment with long correlation times in Sec.~\ref{secadiabatic}. These two limits give very different answers to the questions of whether bet-hedging is desirable, or whether generalist phenotypes can be optimal.
To study the intermediate regime between these two extremes, we first start by presenting the results of a numerical study, based on the recursion equation \ceqref{eqNrecursion}.
We apply the numerical approach described in Ref. \cite{Mayer2016}. 
In short, we approximate the long-term growth rate numerically by simulating for a large number of generations, and then use a derivative-free global optimization algorithm to roughly find the global optimum.
In practice, we focus on two-state environments, which we characterize by their characteristic time scale, $t_c$, defined by $e^{-1/t_c}=1-p(1|2)-p(2|1)$ and the probability of being in state 2, $p(x=2)$.
The numerical results show how the two limits are connected for the case without (Fig.~\ref{figphasesinnate}) and with a generalist phenotype (Fig.~\ref{fig3state}).
\changed{In temporally correlated strategies, phenotype frequencies vary with the environmental history. To represent strategies in a simple way that generalizes the case of memoryless strategies, we define $\pi(\sigma)$ as the steady-state frequency of phenotype $\sigma$ in a lineage, $\sum_{\sigma'}\pi(\sigma|\sigma')\pi(\sigma')=\pi(\sigma)$.}
Consistent with results in the adiabatic limit, for large $t_c$ \changed{switching strategies} dominate across the range of environmental frequencies (Figs.~\ref{figphasesinnate} and \ref{fig3state}): $\forall \sigma, \pi^\star(\sigma)<1$.
In the case where there is an intermediate, generalist phenotype ($\sigma = 3$) the switching takes place \changed{primarily} between the specialist types: \changed{$\pi^\star(\sigma=3) \ll 1$} for large $t_c$ (Fig.~\ref{fig3state}), consistent with the argument that \changed{specialized phenotypes are} optimal in the adiabatic limit (Sec.~\ref{secadiabatic}).
The transition to a regime where non-switching strategies are optimal happens when the temporal correlations of the environment are of the order of the generation time, $t_c \sim 1$. \changed{In this regime, all three phenotypes (two specialists and one generalist) may co-exist in the optimal strategy, $\forall \sigma,\pi^*(\sigma)>0$. Recall that such mixtures involving more phenotypes than distinct environments are suboptimal in memoryless environments, $t_c=0$, as deduced from the graphical construction (see Sec.~\ref{seciidtransitions}).}

\subsection{An analytical result for intermediate timescales}
\label{seccorrana}

We present here an approach to derive analytically the boundaries between optimal switching and non-switching strategies in correlated environments. The approach is based on an expansion at small switching rates of the Master equation of the joint environmental and population switching process near the transition boundary.

We first rewrite \ceqref{eqNrecursion} as a recursion equation for the fraction of the population in each state $n_t(\B \sigma) = N_t(\B \sigma) / N_t$ with $N_t = \sum_{\B \sigma} N_t(\B \sigma)$,
\begin{equation}
    n_{t+1}(\B \sigma) = \frac{1}{Z_t} f(\B \sigma, \B x_t) \sum_{\B \sigma'} \pi(\B \sigma|\B \sigma') n_t(\B \sigma'),
\end{equation}
where $Z_t$ is a normalization constant enforcing $\sum_{\B \sigma} n_t(\B \sigma) = 1$.
Since $N_T = N_0 \prod_{t=0}^T Z_t$ the long-term growth rate given by \ceqref{eqLambdadef} becomes $\Lambda = \lim_{T\rightarrow\infty} \frac{1}{T} \sum_{t=0}^T \ln Z_t$.

For simplicity, we consider a two-state model again.
We introduce the simplified notations $\pi(1 | 2) = \pi_{12}$, $\pi(2 | 1) = \pi_{21}$ for the type switching rates, $p(1|2) = p_{12}$, $p(2|1) = p_{21}$ for the environment switching rates, and denote $n_t(2) = n_t$, and $x_t(2) = x_t$. We use the same convention as in \ceqref{eqselectioncoefficients}, $f(1,1)=f(2,2)=1$, $w(2,1)=w_1$ and $w(1,2)=w_2$.
This allows us to rewrite the recursion equation as 
\begin{equation} \label{eqrecn}
    n_{t+1} = \frac{1}{Z_t} w_1^{1-x_t} \left(n_t (1-\pi_{21}) + (1-n_t) \pi_{12}\right)
\end{equation}
with
\begin{equation}
    Z_t = n_t w_1^{1-x_t} + (1-n_t) w_2^{ x_t}.
\end{equation}

To analyze the transition from an optimal strategy where all individuals have phenotype 1 to a strategy with some switching to the other phenotype, we need to know
whether a small $\pi_{12}$ is better than $\pi_{12} = 0$ -- if that is the case, then \changed{switching} is advantageous.
We thus consider $\pi_{12} \ll 1$ and $n_t \ll 1$.
The recursion \ceqref{eqrecn} becomes at leading order in $n_t$ and $\pi_{12}$:
\begin{equation}\label{eq:rtp0}
    n_{t+1} = \pi_{12} + (1-\pi_{21}) w_1^{1-x_t} w_2^{ x_t} n_t.
\end{equation}
$\ln Z_t$ can also be expanded:
\begin{equation}
    \begin{split}
    \ln Z_t&=x_t \ln w_2 - n_t + n_t w_1^{1-x_t} w_2^{ x_t}\\
    &=x_t  \ln w_2 +n_t(w_1-1) + x_t r_t (w_2^{-1}-w_1).
    \end{split}
\end{equation}
Over long times the joint environmental-population process is ergodic. The long-term growth rate is thus given as $\<\ln Z\>$, where $\langle.\rangle$ indicates an average over the steady state distribution of $x, n$.
No \changed{switching} ($n=0$) gives a long-term growth rate of $ \<x\>\ln w_2$ 
 Thus the difference in long-term growth rate between stochastic switching and the single-phenotype strategy is
\begin{equation} \label{eqdLambda}
    \Delta \Lambda = \<n\> (w_2^{-1} - w_1) \left(\frac{\<x n\>}{\<n\>} - \frac{1-w_1}{w_2^{-1} - w_1} \right),
\end{equation}
which shows that stochastic switching is advantageous if
\begin{equation}\label{eqcond}
    \frac{\<x n\>}{\<n\>} > \frac{1-w_1}{w_2^{-1} - w_1}.
\end{equation}
We can identify the right-hand side of this equation with the lower bound environmental frequency $p_2^{\rm lb}$ in uncorrelated environments defined in \ceqref{eqplb}.
When there is no memory, $n$ and $x$ are uncorrelated, the left-hand side reduces to $\<x\> = p(x=2)$ and we recover the result of \ceqref{eqpioptinnate}.
If there is memory, then $n$ and $x$ are positively correlated through the effects of selection on the population composition, which increases the fraction on the left-hand side. This leads us to a first important conclusion: switching is favored over non-switching strategies under a wider range of environmental parameters in the presence of temporal autocorrelation.

We go further and calculate analytically the left-hand side of \ceqref{eqcond} at the transition.
Some algebra shows that $\rho_{1,t}=\<(1-x_t) n_t\>$ and $\rho_{2,t}=\<x_t n_t\>$ satisfy the recursion
\begin{align}
    \rho_{1,t+1}=&p_{21} [\pi_{12} p_2 + (1-\pi_{21})w_2^{-1}\rho_{2,t}]\nonumber\\
    +&(1-p_{12})[\pi_{12} (1-p_2)+(1-\pi_{21})w_2\rho_{1,t}],\\
    \rho_{2,t+1}=&(1-p_{21})[\pi_{12} p_2 + (1-\pi_{21})w_2^{-1}\rho_{2,t}]\nonumber\\
+&p_{12}[\pi_{12}(1-p_2)+(1-\pi_{21})w_2\rho_{1,t}],
\end{align}
where we use the short-hand notation $p_2 = p_{21}/(p_{12}+p_{21}) = \langle x \rangle$ for the average fraction of generations the environment is in state $2$.
Therefore at steady state, $\rho_{\sigma,t}=\rho_{\sigma,t+1}=\rho_\sigma$, we have
\begin{equation}\label{eqratio}
    \begin{split}
        \frac{\<x n\>}{\<n\>} &= \frac{\rho_2}{\rho_2+\rho_1} \\
        &= \frac{p_2 \left[1 - (1-\pi_{21}) e^{-1/t_c} w_1 \right]}{1 - (1-\pi_{21}) e^{-1/t_c}  w_1 \left[(1-p_2)  w_1w_2 + p_2 \right]},
    \end{split}
\end{equation}
where we recall that $e^{-1/t_c} = 1-p(1 | 2)-p(2 | 1)$ quantifies memory in the environment.
The expression in \eqref{eqratio} is a decreasing function of $\pi_{21}$ so its maximum is achieved in the limit of $\pi_{21}$ going to zero.
Setting $\pi_{21} = 0$ in \ceqref{eqratio} and plugging the result into \ceqref{eqcond}, we obtain the condition needed for optimal switching to outperform always being in state $\sigma=1$:
\begin{equation} \label{eqcorrlb}
    p_2 > \frac{(1-w_1)(1-e^{-1/t_c} w_2^{-1})}{(w_2^{-1}-w_1)(1-e^{-1/t_c})}.
\end{equation}
The second transition, between optimal switching and always being in state $\sigma=2$, is given by the replacements $w_1\to w_2$, $w_2\to w_1$, $p_2 \to 1-p_2$, yielding the condition:
\begin{equation} \label{eqcorrub}
    p_2 < \frac{(w_1^{-1}-1)(1-e^{-1/t_c} w_2)}{(w_1^{-1}-w_2)(1-e^{-1/t_c})}.
\end{equation}

These transitions reduce to \eqref{eqpioptinnate} in the limit of no environment memory, $t_c=0$.
The transition curves reach $p_2=0$ and $p_2=1$ at $t_c = -1/\ln(w_2)$ and $t_c = -1/\ln(w_1)$, respectively.
The resulting phase diagram is shown in Fig.~\ref{figphasesinnate} along with a numerical optimization, which confirms the results.

\begin{figure}
    \begin{center}
    \includegraphics{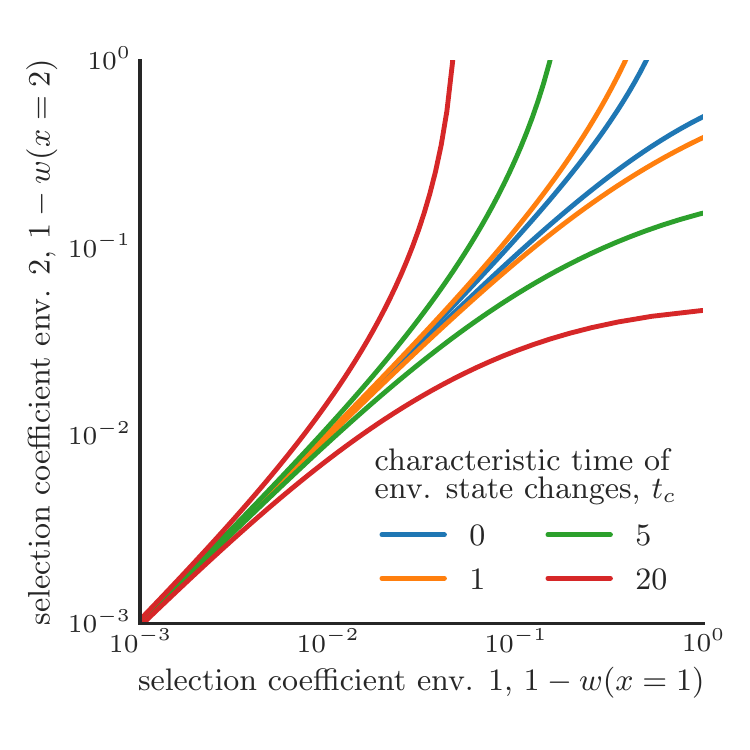}
        \caption{{\bf Environmental correlations increase the range of fitness landscapes for which switching strategies are optimal.}
        Region where switching is optimal (in between colored lines) as a function of environmental correlation time. Two state environment as in Fig.~\ref{figiidbethedging} with symmetric environmental frequencies, $p_2 = p_1 = 0.5$. Selection coefficient $s(x)$ quantifies how much the best adapted phenotype to environment $x$ outperforms the suboptimal phenotype for that environment.
      \label{figasymbhcorr}}
    \end{center}
\end{figure}

The analytical results show that temporal correlations in the environment favor the evolution of stochastic switching.
We can compare to the case of uncorrelated environments considered in Fig.~\ref{figiidbethedging}.
While switching is only optimal in uncorrelated environments if selection is strong in both environments (blue line in Fig.~\ref{figasymbhcorr}), temporally-correlated environments make it optimal for smaller or asymmetric selection (e.g. red line in Fig.~\ref{figasymbhcorr}). We may interpret this broadening of the range where switching is optimal by noting that, in correlated environments, switching does not just contribute to bed-hedging but also to adaptively tracking the state of the environment.

\subsection{Continuous time limit}

Lastly, we discuss the continuous time limit of \ceqref{eqNrecursion} where our results take a simple form. The limit is obtained by rescaling the switching rates, growth rates, and times by $\delta t$, $p(x|x') \to p(x|x') \delta t$ for $x \neq x'$, $\pi(\sigma|\sigma') \to \pi(\sigma|\sigma') \delta t$ for $\sigma \neq \sigma'$, and $\ln[f(\sigma, x)] \to m(\sigma, x) \delta t$, $t \to t/\delta t$, $t_c \to t_c / \delta t$, \changed{$\ln w_x \to \ln w_x/\delta_t$, $\Lambda \to \Lambda/\delta t$} and sending $\delta t \to 0$, which yields
\begin{equation}
    \frac{\ud \B N}{\ud t}  = \B A^{x(t)} \B N(t),
\end{equation}
where $A^{x(t)}_{\sigma, \sigma'} =  m(\sigma, x(t)) \delta_{\sigma, \sigma'} + \pi(\sigma|\sigma')$.

We can take the limit of the results obtained in Sec.~\ref{seccorrana} to see how temporal autocorrelation influences the results in this case.
From \ceqref{eqcorrlb} we obtain
\begin{equation} \label{eqtransition1}
    p_2 > \frac{1+t_c \ln w_2}{1 + \ln w_2/\ln w_1},
\end{equation}
and from \ceqref{eqcorrub}
\begin{equation} \label{eqtransition2}
    p_2 < \frac{1-t_c \ln w_2}{1 + \ln w_2/\ln w_1}.
\end{equation}
The formulas are simpler and notably linear in the correlation time $t_c$.
The range of environmental frequencies for which stochastic switching is optimal thus grows linearly with the environmental correlation time scale $t_c$, as $-2t_c\ln w_1\ln w_2/\ln (w_1w_2)$.

\begin{figure}
\begin{center}
\includegraphics{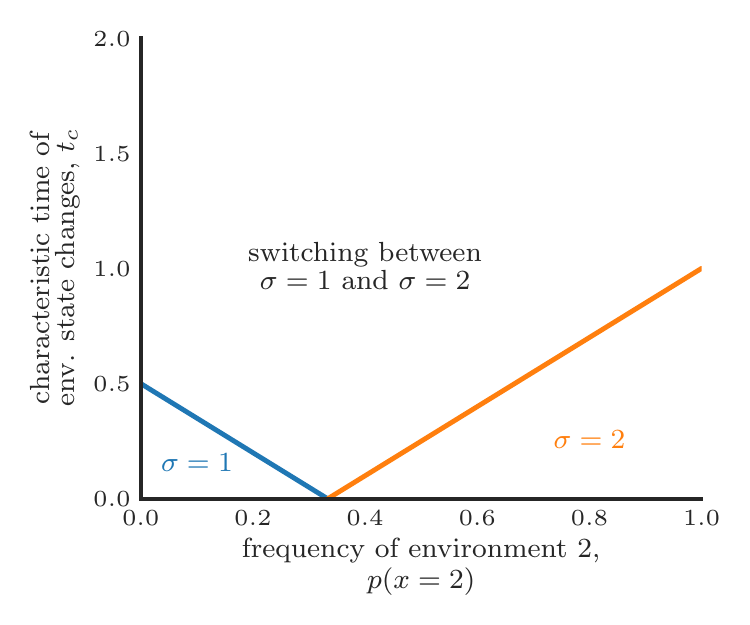}
    \caption{{\bf Phase diagram in the continuous time limit.} $t_c$ is the environment correlation time, and $p$ is the fraction of the time the environment is in state 2. On the left of the \changed{blue line, the optimal solution is for the population to have single phenotype $\sigma=1$. On the right of the red line, the optimal solution is to have single phenotype $\sigma=2$. In between the optimal solution is to switch between both phenotypes. The blue transition line reaches $p=0$ at $t_c = -1/\ln w_1$, while the red transition reaches $p=1$ at $t_c=-1/\ln w_2$.} The two transitions meet at $p=\ln w_1/\ln (w_1w_2)$ (dashed line). Parameters: $\ln w_2=-2$ and $\ln w_1=-1$.
    \label{figphasescontinuous}}
\end{center}
\end{figure}

The point $p_2=0$ is reached by the first transition \ceqref{eqtransition1} from the non-switching to switching regime for $t_c=-1/\ln w_2$, and the point $p_2=1$ reached by the second transition for $t_c=-1/\ln w_1$ \ceqref{eqtransition2}, 
as in the discrete time case. In the limit of no environmental memory, $t_c=0$, the two transitions are at the same point $\ln w_1/\ln (w_1w_2)$: this means that bet-hedging is never advantageous and the transition is from one single-phenotype strategy to the other.
This is in contrast with the solution in discrete time, where there always is a window in which bet-hedging is favored, regardless of the environmental memory.
Since in any finite time interval, the environment cycles through all its states, the population effectively only sees the mean environment.
The long-term growth rate in continuous time is thus given by
\begin{equation} \label{eqLambdalin}
    \Lambda = \sum_x p(x) \sum_\sigma f(\sigma, x) \pi(\sigma)
\end{equation}
which is a linear function in $\pi(\sigma)$.
$\Lambda$ is optimized by putting all weight on the phenotype $\sigma$ with largest average fitness $\sum_x p(x) f(\sigma, x)$.
Thus no switching strategies can be optimal and the optimal strategy always consists of a single phenotype.

\section{Discussion}

Our results provide a unified view of transitions between optimal adaptive strategies in randomly fluctuating environments.
By revisiting the fitness set representation of Levins \cite{Levins1968}, valid  for temporally uncorrelated environments, we presented a graphical method, supplemented by analytical calculations, to determine the transitions between bet-hedging and single-phenotype strategies, as well as between specialist and generalist phenotypes (Fig.~\ref{figtransitions}), generalizing previous results \cite{Cohen1966,Levins1968,Haccou1995,Cover2005,Rivoire2011,Donaldson-Matasci2008}.
Extending the method to phenotypes constrained by a trade-off function, we constructed graphically and calculated analytically the transitions between optimal strategies of diversification and adaptability in a simple model of evolution of immunity \cite{Mayer2016} (Fig.~\ref{figimmune}).

As noticed in previous studies, temporal correlations in the environmental conditions influences the choice of optimal adaptation strategies \cite{Ratcliff2015,Botero2015,Skanata2016}. The intermediate timescale regime, where the environmental correlation time is of the same order as the generation time, has been notoriously difficult to handle analytically. Here, we presented an analytical approach to show how temporal correlations in environments can be exploited by switching strategies that keep some memory of previous phenotypes. Our results show that temporal correlations broaden the range of selective pressures for which a switching strategy is better than a single-phenotype one.
Everything else being equal,
switching strategies are thus more favorable in correlated environments than in uncorrelated environments.
To our knowledge, only one other analytical approach is available to analyze optimal strategies in correlated environments \cite{Skanata2016}.

The results are independent of mechanisms, which may take different forms. For instance, one mechanism to achieve a generalist phenotype is through {\it plasticity}, i.e., a generalist phenotype may partly or totally be induced by the environmental condition. In our approach, however, only the value of the replication rate $f(\sigma, x)$ in environmental condition $x$ given the inherited type $\sigma$ matters, not the process by which it is achieved. Only when the induced phenotype may be transmitted to the next generation, as for instance with the Lamarckian CRISPR-like strategy of  \cite{Mayer2016}, does the distinction between inherited and induced phenotype, and therefore the concept of plasticity, become relevant.

Possible extensions of our results include the
 influence of non-random environmental changes, such as periodic environments \cite{Gaal2010,Carja2014,Botero2015,Skanata2016}, constraints on relative switching rates \cite{Salathe2009,Gaal2010,Patra2015}, active sensing mechanisms \cite{Kussell2005} and heritable plasticity \cite{Rivoire2014,Mayer2016}, or finite population size effects \cite{Xue2017}.
Some of these factors are known to lead to transitions between adaptive strategies, e.g. the variability of environmental durations \cite{Skanata2016}, or cause the transitions to become discontinuous, e.g. when switching rates are constrained to be independent of phenotype \cite{Salathe2009,Gaal2010,Patra2015}.

{\bf Acknowledgements.}  
We thank B. Xue and O. Carja for helpful discussions. The work was supported by grant ERCStG n. 306312.

\appendix
\setcounter{figure}{0}
\makeatletter 
\renewcommand{\thefigure}{S\@arabic\c@figure}
\makeatother
\section{Optimal strategies by mapping to unit simplex}
\label{appmapping}

\begin{figure*}
    \begin{center}
        \includegraphics{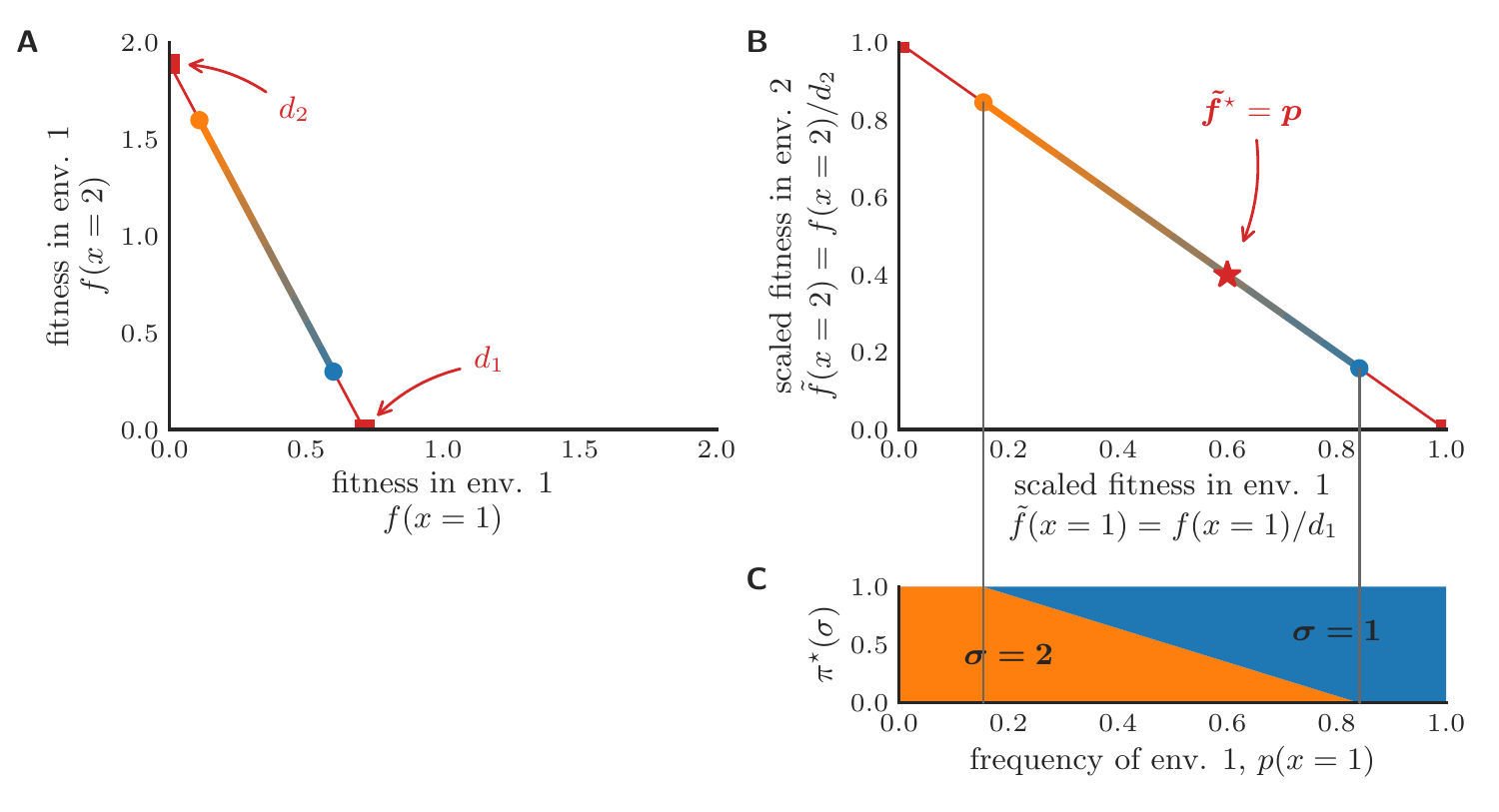}
        \caption{{\bf Mapping of the problem to the unit simplex helps optimizing long-term growth rate graphically.} 
            To determine the best strategy using two phenotypes (blue/orange dots) and their mixtures (colored line) we rescale the original fitnesses (A) such that the sum of fitnesses is constant (B).
            To do so fitnesses are rescaled by dividing through the intercepts (red squares) of the line passing through the two points with the axes (red line). 
            In the scaled fitnesses the optimal strategy has fitness vector $\B{\tilde f}^\star = \B p$ (red star), which can be be mapped back to the original problem by reverting the rescaling.
            Where the so-determined fitnesses lie between the fitnesses of the two phenotypes the optimal strategy switches between both phenotypes with frequencies relative to how far the optimum is from the two phenotypes.
            If the optimal rescaled fitness lies outside the achievable range of fitnesses using the closest phenotype is optimal.
            (C) Optimal mixture of the two phenotypes as a function of the frequency of environmental state $1$. 
    \label{figunitsimplex}}
    \end{center}
\end{figure*}

If the sum of fitnesses of a phenotype over environments $f(\sigma) = \sum_x f(\sigma, x)$ is constant for all phenotypes, then any mixture will also have a constant sum of fitnesses. The normalization constraint on $\B \pi$ then translates into an equivalent constraint on $\B f$.
The solution of the optimization problem \changed{in its fitness form} is then particularly simple \cite{Haccou1995,Donaldson-Matasci2008}.
Therefore, where possible, it is worthwhile to map the optimization problem to this simpler case by a rescaling of fitnesses in different environments.
Here we show how to perform the rescaling and the conditions under which it is possible.
Fig.~\ref{figunitsimplex} illustrates such a mapping in a simple case with two environmental states.

The optimization problem is invariant with respect to additions of terms that are constant with respect to the variables over which one optimizes.
Specifically, we can add the term $\sum_x p(x) \log c(x)$ to Eq.~\eqref{eqLambdalog} with all positive $c(x)$, which is constant with respect to $\B \pi$. 
This gives us a new optimization problem with the objective function
$
     \tilde \Lambda = \sum_x p(x) \log [f(x) c(x)] = \sum_x p(x) \log \tilde f(x),
$
in terms of the rescaled fitnesses $\tilde f(x) = \sum_\sigma \pi(\sigma) \tilde f(\sigma, x)$ and $\tilde f(\sigma, x) = f(\sigma, x) c(x)$.
The equivalence of these problems shows that a rescaling of the axes of fitness space does not change the optimal adaptation strategy.

We can now try and use this rescaling to make the sum of scaled fitnesses a constant, which we chose to be $1$ without restriction of generality.
This means we aim to chose $c(x)$, such that
$
    \sum_x \tilde f(\sigma, x) = \sum_x f(\sigma, x) c(x) = 1
$
holds for all $\sigma$.
In matrix-vector notation we can represent these conditions as the systems of equation 
\begin{equation} \label{eqcondc}
    \B F \B c = \B 1,
\end{equation}
where $\B 1 = (1, 1, ..., 1)^T$ is the vector of all ones and $\B F$ the matrix of phenotype fitness profiles with entries $F_{\sigma, x} = f(\sigma, x)$.
Eq.~\eqref{eqcondc} requires that the scalar products of $\B c$ with the row vectors of $\B f$ (the phenotypes fitness profiles) are equal to $1$ for all rows.
The vector $\B c$ thus is a normal vector to the hyperplane spanned by the fitness profiles.
The mapping is therefore only possible if a hyperplane passing through all fitness profiles exists.
The intercept $d_x$ of the hyperplane with the $x$ axis is given by $\B c (d_x \B e_x) = 1 \Leftrightarrow d_x = 1/c_x$, where $\B e_x$ is the $x$-th unit vector.
Eq.~\eqref{eqcondc} thus specificies that we should rescale fitnesses by dividing through these intercepts to achieve our goal of mapping the problem to the unit simplex.
The positivity of the scaling constants $c(x)$ puts further requirements on $\B F$ for the mapping to work: geometrically, all intercepts need to be positive, or algebraically, the inverse of the fitness matrix needs to have all positive row sums.
In the case where $\B F$ is an invertible matrix fitnesses should be rescaled using
$
    \B c = \B F^{-1} \B 1.
$
If the scaling is possible then in the scaled variables the normalization constraint on $\B \pi$ leads to a normalization constraint on $\B{\tilde f}$.

We can derive the optimal fitness profile using the Lagrange formalism.
The Lagrangian of the optimization problem is
\begin{equation}
    \mathcal{L} = \sum_x p(x) \ln \tilde f(x) - \lambda \left(\sum_x \tilde f(x) - 1\right),
\end{equation}
where we have assumed that the optimum is in the interior of $D_{\tilde f}$, i.e. none of the non-negativity constraints on elements $\B \pi$ are active.
Taking the derivative with respect to $\tilde f(x)$ and setting it to zero yields $\tilde f^\star(x) = p(x)/\lambda$.
As we have rescaled fitnesses such that the sum of fitnesses scale to $1$ the Lagrange multiplier is $\lambda = 1$.
In the rescaled variables the optimal strategy thus allocates fitness to each environment proportional to its frequency:
\begin{equation}
    \B{\tilde f}^\star = \B p.
\end{equation}
From the optimum in the rescaled variables the optimum in the original variables can be obtained by reversing the scaling $f^\star_i = \tilde f^\star_i/c_i$.

Due to the non-negativity constraints on $\pi(\sigma)$, which we have neglected so far in the discussion, only a subset of the unit simplex is accessible if there are no phenotypes that are not completely specialized to the different environments.
Where the unconstrained solution lies outside the feasible region a value on the boundary of the fitness set is constrained optimum instead.
The fitness allocation among the remaining unconstrained directions still is proportional to the frequency of the respective environments.

\section{Analytical results on optimal immune strategies in uncorrelated environments}
\label{appiid}

\subsection{Optimization problem}
The cost function of the optimization problem is the long-term population growth rate, which depends on the environmental statistics and the chosen strategy.
The long-term growth rate in uncorrelated environments for a given $p, f_{\rm base}, f_{\rm inf}, f_{\rm def}(f_{\rm con})$ is given by \ceqref{eqLambdadirect}, which we recall here
\begin{equation} \label{eqLambda}
\begin{split}
    \Lambda(\pi, f_{\rm con}) &= p \ln[\pi f_{\rm def} + (1-\pi) f_{\rm inf}]  \\
&+ (1-p) \ln [\pi f_{\rm con} + (1-\pi) f_{\rm base}].
\end{split}
\end{equation}
To find the optimal strategy we need to solve the following optimization problem
\begin{equation}
\begin{aligned}
    &\underset{\pi, f_{\rm con}}{\text{maximize}} &&\Lambda(\pi, f_{\rm con}) \\
    &\text{subject to}  &&0 \leq \pi \leq 1 \\
    &                   &&f_{\rm con}^{min} \leq f_{\rm con} \leq f_{\rm con}^{max}
\end{aligned}.
\end{equation}
The optimization consists in finding the (global) maximum of a two-variable objective function subject to bound constraints on both variables.
In the following derivations we make use of the ordering of the costs in the non-trivial case $f_{\rm base} \geq f_{\rm con}, f_{\rm def} > f_{\rm inf}$ and of the Pareto condition on the trade-off line $f_{\rm def}'(f_{\rm con}) < 0$.

This problem can be solved numerically, but as is shown in the following a lot of information is available from a purely analytical treatment of the optimization problem.
The Karush-Kuhn-Tucker conditions give necessary conditions for local optimality of a point $\pi^\star, f_{\rm con}^\star$.
For bound constrained problems these conditions boil down to the statement that the partial derivative of the objective function with respect to either variable needs to be \cite{Boyd2004}:
zero if the variable is in the interior of its feasible interval, negative if the variable is at the lower end of its feasible domain, and positive if the variable is at the upper end of its feasible domain.
Expressed in equations the necessary conditions for $\pi^\star, f_{\rm con}^\star$ to be locally optimal is that
\begin{equation} \label{eqoptcondpi}
    \partial_{\pi} \Lambda(\pi^\star, f_{\rm con}^\star)
\begin{cases}
    \leq 0, \, \text{if} \, \pi^\star = 0 \\
    \geq 0, \, \text{if} \, \pi^\star = 1 \\
    = 0, \, \text{otherwise}
\end{cases}
\end{equation}
and that
\begin{equation} \label{eqoptcondmu1}
    \partial_{f_{\rm con}} \Lambda(\pi^\star, f_{\rm con}^\star)
\begin{cases}
\leq 0, \, f_{\rm con}^\star = 0 \\
\geq 0, \, f_{\rm con}^\star = 1 \\
= 0, \, \text{otherwise}
\end{cases}.
\end{equation}
The conditions provide only necessary but not sufficient conditions for local optimality.
A condition ensuring sufficiency is that the Hessian at the optimum constricted to the feasible directions is negative definite.

\subsection{Derivatives of the cost function}
The optimality conditions derived in the previous subsection involve the derivatives of the cost function, which can be obtained using simple algebra and which we give below.
The derivative of the cost function with respect to $\pi$ is given by
\begin{align}
    \partial_{\pi} \Lambda &= \frac{p (f_{\rm def}-f_{\rm inf})}{f_{\rm inf} (1-\pi)+f_{\rm def} \pi}
                             - \frac{(1-p)(f_{\rm base}-f_{\rm con})}{f_{\rm base} (1-\pi)+f_{\rm con} \pi}  \\
    \partial_{f_{\rm con}} \Lambda &= \pi \left[\frac{p f_{\rm def}'}{f_{\rm inf} (1-\pi)+f_{\rm def}\pi}
                             + \frac{(1-p)}{f_{\rm base} (1-\pi)+f_{\rm con} \pi} \right].
\end{align}
For sufficiency we also need to consider the second derivatives of the cost function:
\begin{widetext}
\begin{align}
    \partial_{\pi}^2 \Lambda &= - \frac{p (f_{\rm def}-f_{\rm inf})^2}{(f_{\rm inf} (1-\pi)+f_{\rm def} \pi)^2}
                             - \frac{(1-p)(f_{\rm base}-f_{\rm con})^2}{(f_{\rm base} (1-\pi)+f_{\rm con} \pi)^2}  \\
    \partial_{f_{\rm con}}^2 \Lambda &= - \pi \left[\frac{p (\pi f_{\rm def}'^2 -(f_{\rm inf} (1-\pi) + \pi f_{\rm def}) f_{\rm def}'')}{(f_{\rm inf} (1-\pi)+f_{\rm def}\pi)^2}
                             + \frac{(1-p) \pi}{(f_{\rm base} (1-\pi)+f_{\rm con} \pi)^2} \right].
\end{align}
\end{widetext}
The second derivative with respect to $\pi$ is always negative which shows that the long-term growth rate is a concave function of $\pi$.
For a fixed value of $f_{\rm con}$ the optimization thus corresponds to a maximization of a concave function and always yields a unique optimum.
The second derivative of the long-term growth rate with respect to $f_{\rm con}$ is also always negative, if $f_{\rm def}'' \leq 0$.
This condition on the trade-off function is fulfilled if individuals might bet hedge in their degree of specialization in environment $1$.
Otherwise the second derivative might be positive for some $p$ and there can thus exist more than one local optimal in the full optimization problem.

A sufficient condition for having a local maximum is the negative definiteness of the Hessian.
As one of its diagonal elements is always negative this is equivalent to showing that the determinant of the Hessian is positive.
The determinant of the Hessian at an interior stationary point can be calculated to be 
\begin{widetext}
\begin{equation}
    \det \nabla^2 \Lambda(\pi^\star, f_{\rm con}^\star)
    = - f_{\rm def}''(f_{\rm con}^\star)
    \frac{(f_{\rm base} - f_{\rm con}^\star)^2 (f_{\rm def}^\star - f_{\rm inf}) \pi }{(f_{\rm base} f_{\rm def}^\star-f_{\rm inf} f_{\rm con}^\star) (f_{\rm base} (1 - \pi) + f_{\rm con}^\star \pi)  (f_{\rm inf} (1 - \pi) + \pi f_{\rm def}^\star)}
    .
\end{equation}
\end{widetext}
It follows that for $f_{\rm con}^\star$ to be optimal for an intermediate $\pi^\star$ the trade-off curve needs to be locally concave $f_{\rm def}''(f_{\rm con}^\star) < 0$.

\subsection{Regions of local optimality for different phases}
The optimality conditions Eqs.~\eqref{eqoptcondpi} and \eqref{eqoptcondmu1} allow for three different cases for $\pi^\star$ and $f_{\rm con}^\star$ each.
This makes for a total of $3 \times 3 = 9$ different combinations. 
For the case $\pi^\star = 0$ the growth rate does not depend on $f_{\rm con}^\star$, so there exists up to seven distinct phases.
Under which conditions are these strategies locally optimal?
In the following we analytically derive the interval of $p$ for which these strategies are optimal.

\subsubsection{Tolerance ($\pi^\star = 0$, arbitrary $f_{\rm con}^\star$)}
The condition of local optimality is $\partial_\pi \Lambda(0, f_{\rm con}^\star) \leq 0$ (see Eq.~\eqref{eqoptcondpi}), which needs to hold for all feasible $f_{\rm con}^\star$. This translates to the condition
$p \leq \frac{f_{\rm inf} (f_{\rm base} - f_{\rm con}^\star)}{f_{\rm base} f_{\rm def}^\star - f_{\rm inf} f_{\rm con}^\star}$.
The condition needs to hold for the $f_{\rm con}^\star$ giving the strictest bound. Tolerance thus is optimal for
\begin{equation}
    p \leq \min_{f_{\rm con}} \frac{f_{\rm inf} (f_{\rm base} - f_{\rm con})}{f_{\rm base} f_{\rm def} - f_{\rm inf} f_{\rm con}} =: p^{(0)},
\end{equation}
i.e. for the rarest pathogens.
If the adaptive strategy comes without constitutive cost ($f_{\rm con}^{min} = 0$), then the tolerance phase disappears ($p^{(0)} = 0$).
Where the phase exists it is followed by one of the bet hedging strategies.

\subsubsection{Innate ($\pi^\star=1, f_{\rm con}^\star=f_{\rm con}^{min}$)}
From \ceqref{eqoptcondpi} the condition of local optimality is $\partial_\pi \Lambda(1, f_{\rm con}^{min}) \geq0$.
This translates to the condition
\begin{equation}
    p \geq \frac{(f_{\rm base} - f_{\rm con}^{min}) f_{\rm def}^{max}}{f_{\rm base}f_{\rm def}^{max}-f_{\rm inf}f_{\rm con}^{min}} =: p^{(i\tilde i)}.
\end{equation}
The second optimality condition \ceqref{eqoptcondmu1} is $\partial_{f_{\rm con}} \Lambda(1, f_{\rm con}^{min}) \geq 0$, leading to
\begin{equation}
    p \geq \frac{f_{\rm def}^{max}}{f_{\rm def}^{max}-f_{\rm con}^{min} f_{\rm def}'(f_{\rm con}^{min})} =: p^{(ip)}.
\end{equation}
Both conditions need to hold at the same time for local optimality so an innate strategy is optimal for
\begin{equation}
    p \geq \max (p^{(ip)}, p^{(i\tilde i)}),
\end{equation}
i.e. for the most frequent pathogens.
Depending on which of the two conditions is more stringent it is followed either by a innate bet hedging strategy or a protoadaptive phase.

\subsubsection{Adaptive ($\pi^\star=1, f_{\rm con}^\star=f_{\rm con}^{max}$)}
Eq.~\eqref{eqoptcondpi} leads to
\begin{equation}
    p \geq \frac{(f_{\rm base} - f_{\rm con}^{max}) f_{\rm def}^{min}}{f_{\rm base}f_{\rm def}^{min}-f_{\rm inf}f_{\rm con}^{max}} =: p^{(a\tilde a)}
\end{equation}
and Eq.~\eqref{eqoptcondmu1} to
\begin{equation}
    p \leq \frac{f_{\rm def}^{min}}{f_{\rm def}^{min}-f_{\rm con}^{max} f_{\rm def}'(f_{\rm con}^{max})} =: p^{(ap)}.
\end{equation}
Taken together an adaptive strategy is optimal for
\begin{equation}
     p^{(a\tilde a)} \leq p \leq p^{(ap)}.
\end{equation}

\subsubsection{Protoadaptive ($\pi^\star=1$, intermediate $f_{\rm con}^\star$)}
Eq.~\eqref{eqoptcondpi} leads to $p \geq \frac{(f_{\rm base} - f_{\rm con}^\star) f_{\rm def}^\star}{f_{\rm base}f_{\rm def}^\star-f_{\rm inf}f_{\rm con}^\star} $ and Eq.~\eqref{eqoptcondmu1} to $p = \frac{f_{\rm def}^\star}{f_{\rm def}^\star-f_{\rm con}^\star f_{\rm def}'(f_{\rm con}^\star)}$.
The two conditions together lead to
\begin{equation} \label{eqfprime}
    f_{\rm def}'(f_{\rm con}^\star) \geq - \frac{f_{\rm def}(f_{\rm con}^\star) - f_{\rm inf} }{f_{\rm base} - f_{\rm con}^\star},
\end{equation}
i.e. the derivative of the trade-off function needs to be more shallow then the derivative of costs of a mixture with the current type. 
As we have an intermediate level of regulation we need to check the second derivative to assure the extremum is a maximum.
As shown in the main text this leads to the condition $\frac{\ud^2 \ln f_{\rm def}}{\ud (\ln f_{\rm con})^2} < 0$. 
If the trade-off function is assumed to be fulfill both conditions everywhere and to be smooth then by the intermediate value theorem there is a $f_{\rm con}^\star$, which is optimal for a $p$ in the region
\begin{equation}
    p^{(ap)}
    \leq p \leq  
    p^{(ip)}
\end{equation}

\subsubsection{innate switching (intermediate $\pi^\star$, $f_{\rm con}^\star = f_{\rm con}^{min}$)}
Eq.~\eqref{eqoptcondpi} leads to
\begin{equation}
    p^{(0\tilde i)} \leq p  \leq p^{(i \tilde i)}
\end{equation}
with
\begin{equation}
    p^{(0\tilde i)} =  \frac{f_{\rm inf} (f_{\rm base} - f_{\rm con}^{min})}{f_{\rm base} f_{\rm def}^{max} - f_{\rm inf} f_{\rm con}^{min}}
\end{equation}
and Eq.~\eqref{eqoptcondmu1} to 
\begin{equation}
    f_{\rm def}'(f_{\rm con}^{min}) \leq - \frac{f_{\rm def}^{max} - f_{\rm inf}}{f_{\rm base} - f_{\rm con}^{min}}
\end{equation}
i.e. the derivative of the trade-off shape needs to be steeper then the line joining the unprotected state. 

\subsubsection{adaptive switching (intermediate $\pi^\star$, $f_{\rm con}^\star = f_{\rm con}^{max}$)}
Eq.~\eqref{eqoptcondpi} leads to
\begin{equation}
     p^{(0\tilde a)} \leq p  \leq p^{(a \tilde a)}
\end{equation}
with
\begin{equation}
    p^{(0\tilde a)} =  \frac{f_{\rm inf} (f_{\rm base} - f_{\rm con}^{max})}{f_{\rm base} f_{\rm def}^{min} - f_{\rm inf} f_{\rm con}^{max}}
\end{equation}
and Eq.~\eqref{eqoptcondmu1} to 
\begin{equation}
    f_{\rm def}'(f_{\rm con}^{min}) \leq - \frac{f_{\rm def}^{min} - f_{\rm inf}}{f_{\rm base} - f_{\rm con}^{max}}.
\end{equation}

\subsubsection{protoadaptive switching (intermediate $\pi^\star, f_{\rm con}^\star$)}
Eq.~\eqref{eqoptcondpi} leads to
\begin{equation}
    p^{(0\tilde p)} \leq p  \leq p^{(p \tilde p)}
\end{equation}
with
\begin{equation}
    p^{(0\tilde p)} =  \frac{f_{\rm inf} (f_{\rm base} - f_{\rm con}^\star)}{f_{\rm base} f_{\rm def}^\star - f_{\rm inf} f_{\rm con}^\star}
\end{equation}
and Eq.~\eqref{eqoptcondmu1} to 
\begin{equation} \label{eqoptcondmu1pbh}
    f_{\rm def}'(f_{\rm con}^\star) = - \frac{f_{\rm def}(f_{\rm con}^\star) - f_{\rm inf} }{f_{\rm base} - f_{\rm con}^\star}.
\end{equation}
The derivative needs to be equal to the slope of the line connecting the fitness profile to the non-protected type.
The sufficiency condition $\det H(\pi^\star, f_{\rm con}^\star) > 0$ leads to
\begin{equation}
    f_{\rm def}''(f_{\rm con}^\star) < 0.
\end{equation}

\section{Derivation of long-term growth rate in the adiabatic limit}
\label{appadiabatic}
The study of the adiabatic limit in which the durations of environmental periods are large relative to the time scales of population composition change goes back to \cite{Kussell2005}. Mathematically the long-term growth rate can be approximated by an eigenvalue perturbation approach. In the following we give a derivation following the notations of \cite{Rivoire2011}. 

The transfer matrix connecting the population composition at successive time points is $\bra{\sigma'}A^{(x)}\ket{\sigma} = f(\sigma', x) \pi(\sigma'|\sigma)$ (in bra-ket notation), which one can decompose as $A^{(x)} = A_0^{(x)} + A_1^{(x)}$ with 
\begin{equation}
    \bra{\sigma'}A_0^{(x)}\ket{\sigma}=\begin{cases} f(\sigma, x) \; \text{if} \; \sigma' = \sigma \\
        0 \; \text{otherwise}.
    \end{cases}
\end{equation}
and 
\begin{equation}
    \bra{\sigma'}A_1^{(x)}\ket{\sigma}=\begin{cases} -f(\sigma, x) (1-\pi(\sigma|\sigma)) \; \text{if} \; \sigma' = \sigma \\
                                                     f(\sigma', x) \pi(\sigma'|\sigma) \; \text{otherwise}.
    \end{cases}
\end{equation}
Using this decomposition we treat $A_1^{(x)}$ as a perturbation to $A_0^{(x)}$ to approximately solve the eigenvalue problem of $A^{(x)}$.
As $A_0$ is diagonal its eigenvalues are $\lambda_{0, \sigma} = f(\sigma, x)$ with corresponding eigenvectors $\ket{\sigma}$, which have all but the $\sigma$-th element set to zero. 
Applying the formulas for the eigenvalues and eigenvectors of the perturbed problem we obtain
\begin{equation}
    \lambda_{\sigma} = f(\sigma, x) \pi(\sigma | \sigma)
\end{equation}
and the corresponding right eigenvectors
\begin{align}
    \ket{\psi_\sigma^{(x)}} &= \ket{\sigma} + \sum_{\sigma' \neq \sigma} \frac{\bra{\sigma'}A_1^{(x)}\ket{\sigma}}{f(\sigma, x) - f(\sigma', x)} \ket{\sigma'} \\
                             &=  \ket{\sigma} + \sum_{\sigma' \neq \sigma} \frac{f(\sigma', x) \pi(\sigma' | \sigma)}{f(\sigma, x) - f(\sigma', x)} \ket{\sigma'}
    .
\end{align}
In order to calculate overlaps we also need to calculate left eigenvectors.
The left eigenvectors of $A_0^{(x)}$ are equal to its right eigenvectors as its a diagonal matrix. The left eigenvectors of the perturbed problem are
\begin{align}
    \bra{\psi_\sigma^{(x)}} &= \bra{\sigma} + \sum_{\sigma' \neq \sigma} \frac{\bra{\sigma'}(A_1^{(x)})^T\ket{\sigma}}{f(\sigma, x) - f(\sigma', x)} \bra{\sigma'} \\
    &= \bra{\sigma} + \sum_{\sigma' \neq \sigma} \frac{\bra{\sigma}A_1^{(x)}\ket{\sigma'}}{f(\sigma, x) - f(\sigma', x)} \bra{\sigma'} \\
    &= \bra{\sigma} + \sum_{\sigma' \neq \sigma} \frac{f(\sigma, x) \pi(\sigma|\sigma')}{f(\sigma, x) - f(\sigma', x)} \bra{\sigma'}.
\end{align}
Let us assume that for every environment $x$ there is a type $\sigma = x$, which provides optimal growth.
The overlap between the largest eigenvectors in environments $x$ and $x'$ is given by
\begin{equation}
    Q(x, x'):=\braket{\psi_x^{(x)}}{\psi_{x'}^{(x')}}   = \pi(x | x')\Gamma(x, x')
\end{equation}
with
\begin{equation}
\Gamma(x, x') =\frac{f(x, x')}{f(x', x') - f(x, x')} + \frac{f(x, x)}{f(x, x) - f(x', x)}
\end{equation}
In the adiabatic limit the long-term growth rate is given by
\begin{align}
    \Lambda =& \sum_x p(x) \ln \lambda_x - \sum_{x, x'; x \neq x'} p(x' | x) p(x) \ln \frac{1}{Q(x, x')} \\
    =& \sum_x p(x) \ln f(x, x) \nonumber \\
&+ \sum_{x, x'} p(x' | x) p(x) \ln  [\pi(x | x') \Gamma(x, x')],
\end{align}
where we have defined $\Gamma(x, x) = 1$.

\changed{
We can write out the sums in the case of an environment switching between two states as
\begin{align}
    \Lambda =&  p(1) \ln f(1, 1) + p(2) \ln f(2, 2)  \nonumber \\
    &+ p(1 | 2) p(2) \ln  [\pi(1 | 2) \Gamma(1, 2)] \nonumber \\
    &+ p(2 | 1) p(1) \ln  [\pi(2 | 1) \Gamma(2, 1)].
\end{align}
To compare the best switching strategies using phenotypes of fitness $\B f$ or $\B{\tilde f}$ we calculate the long-term growth rate difference
\begin{align}
    \Delta \Lambda =&  (1-p_2) \ln \frac{f(1, 1)}{\tilde f(1, 1)} + p_2 \frac{f(2, 2)}{\tilde f(2, 2)}  \nonumber \\
                    &+ (1-e^{-1/t_c})p_2(1-p_2) \ln \frac{\Gamma(1, 2)\Gamma(2, 1)}{\tilde \Gamma(1, 2)\tilde \Gamma(2, 1)},
\end{align}
where we have used short-hand notations for the environmental switching frequencies as introduced in the text.
Setting $\Delta \Lambda = 0$ we can solve for the transition line between the two sets of phenotypes,
\begin{equation} \label{eqtcspecialists}
    e^{-1/t_c} = 1-\frac{(1-p_2) \ln \frac{\tilde f(1, 1)}{f(1, 1)} + p_2 \ln\frac{\tilde f(2, 2)}{f(2, 2)}}{(1-p_2)p_2 \ln \frac{\Gamma(1, 2)\Gamma(2, 1)}{\tilde \Gamma(1, 2)\tilde \Gamma(2, 1)}}.
\end{equation}
Such an analysis can be applied to the case where a generalist phenotype is on the Pareto frontier to find when switching only uses specialists (Fig.~\ref{fig3state}). To do so we compare the growth rate of switching between the specialist phenotypes $\sigma=1$ and $\sigma=2$ to the growth rates of switching between one of the specialists and the generalist $\sigma=3$. \eqref{eqtcspecialists} then gives an approximate result for the time scale of environmental correlations above which switching only involves specialists.
}

\end{document}